# Low-Latitude Aurorae during the Extreme Space Weather Events in 1859

**Hisashi Hayakawa[1*, 2], Yusuke Ebihara[3, 4], David P. Hand[5], Satoshi Hayakawa[6], Sandeep Kumar[7], Shyamoli Mukherjee[7], and B.Veenadhari[7]**

[1] Graduate School of Letters, Osaka University, Toyonaka, 5600043, Japan (JSPS Research Fellow)

[2] Science and Technology Facilities Council, RAL Space, Rutherford Appleton Laboratory, Harwell Campus, Didcot, OX11 0QX, UK

[3] Research Institute for Sustainable Humanosphere, Kyoto University, Uji, 6100011, Japan

[4] Unit of Synergetic Studies for Space, Kyoto University, Kyoto, 6068306, Japan

[5] Independent Researcher, Washington DC, 20017, the United States

[6] Faculty of Engineering, The University of Tokyo, 1130033, Tokyo, Japan.

[7] Indian Institute of Geomagnetism, Plot 5, Sector 18, New Panvel (West), Navi Mumbai, 410218, India.

Corresponding author: Hisashi Hayakawa (hayakawa@kwasan.kyoto-u.ac.jp)

**Abstract**

The Carrington storm (September 1/2, 1859) is one of the largest magnetic storms ever observed and it has caused global auroral displays in low-latitude areas, together with a series of multiple magnetic storms during August 28 and September 4, 1859. In this study, we revisit contemporary auroral observation records to extract information on their elevation angle, color, and direction to investigate this stormy interval in detail. We first examine their equatorward boundary of "auroral emission with multiple colors" based on descriptions of elevation angle and color. We find that their locations were 36.5° ILAT on August 28/29 and 32.7° ILAT on September 1/2, suggesting that trapped electrons moved to, at least, $L\sim1.55$ and $L\sim1.41$, respectively. The equatorward boundary of "purely red emission" was likely located at 30.8° ILAT on September 1/2. If "purely red emission" was a stable auroral red arc, it would suggest that trapped protons moved to, at least, $L \sim 1.36$. This reconstruction with observed auroral emission regions provides conservative estimations of magnetic storm intensities. We compare the auroral records with magnetic observations. We confirm that multiple magnetic storms occurred during this stormy interval, and that the equatorward expansion of the auroral oval is consistent with the timing of magnetic disturbances. It is possible that the August 28/29 interplanetary coronal mass ejections (ICMEs) cleared out the interplanetary medium, making the ICMEs for the Carrington storm on September 1/2 more geoeffective.

**1 Introduction**

It is known that extreme interplanetary coronal mass ejections (ICMEs) released from sunspots can cause severe magnetic storms, especially when they have southward magnetic fields (e.g., Tsurutani et al., 1992, 2008; Gonzalez et al., 1994; Daglis, 2000, 2004; Daglis and Akasofu, 2004; Willis & Stephenson, 2001; Willis et al., 2005; Echer et al., 2008b; Vaquero et al., 2008; Vaquero & Vazquez, 2009; Schrijver et al., 2012; Odenwald, 2015; Lakhina & Tsurutani, 2016; Hayakawa et al., 2017c; Usoskin, 2017; Takahashi and Shibata, 2017; Riley et al., 2018). During magnetic storms, the horizontal component of geomagnetic fields decreases at low and middle latitudes (Gonzalez and Tsurutani, 1987; Gonzalez et al., 1994). Among the magnetic observations over approximately the past 1.5 centuries, the largest magnetic storm ever observed is considered the Carrington storm in 1859 (Chapman & Bartels, 1940; Jones, 1955; Chapman, 1957; Mayaud, 1980; Tsurutani et al., 2003; Cliver & Svalgaard, 2004; Lakhina and Tsurutani, 2016, 2017). Recent studies suggest evidence of several intense magnetic storms in the coverage of magnetic observations such as those in 1872 (Silverman, 1995, 2006, 2008; Silverman & Cliver, 2001; Vaquero et al., 2008; Cliver & Dietrich, 2013; Viljanen et al., 2014; Lefèvre et al., 2016; Lockwood et al., 2017; Lakhina and Tsurutani, 2017; Love, 2018; Hayakawa et al., 2018b; Riley et al., 2018), satellite observations of a near miss extreme ICME in 2012 (Baker et al., 2013; Liu et al., 2014), and historical evidence before the coverage of magnetic observations (Willis et al., 1996; Ebihara et al., 2017; Hayakawa et al., 2016b, 2017a, 2017b, 2017c, 2018a).

On September 1, Carrington (1859) and Hodgson (1859) witnessed a white light flare within a sunspot group as large as 2300~3000 msh (millionths of solar hemisphere) (e.g. Cliver & Keer, 2012; Hayakawa et al., 2016a), just before the maximum of the solar cycle 10 in 1860 (Clette et al., 2014; Clette and Lefevre, 2016; Svalgaard and Schatten, 2016). This flare is estimated to be X45±5 in terms of SXR class based on the amplitude of magnetic crochet and considered one of the most extreme flares in observational history (Boteler, 2006; Cliver and Dietrich, 2013). On the following day (September 1/2), the ICMEs released from this active region brought intense magnetic storms with a maximum negative intensity of ~1600 nT at Colaba (Tsurutani et al., 2003; Nevanlinna, 2004, 2006, 2008; Viljanen et al., 2014; Kumar et al., 2015; Lakhina and Tsurutani, 2016, 2017). Great auroral displays in low-latitude areas were reported at observation sites down to 22–23° magnetic latitude (hereafter, MLAT), as shown in Figure 1 (Kimball, 1960; Tsurutani et al., 2003; Cliver & Svalgaard, 2004; Cliver & Dietrich, 2013; Hayakawa et al., 2016a; Lakhina & Tsurutani, 2016, 2017). In addition to this storm, multiple magnetic storms occurred during the interval between August 28 to September 4, 1859 (Kimball, 1960; Green et al., 2006; Green & Boardsen, 2006; Hayakawa et al., 2016a; Lakhina & Tsurutani, 2017), resulting from multiple flaring from the solar active region that could produce the multiple ICMEs and multiple sheaths, as is usually the case with extreme events (Mannucci et al., 2005; Willis et al., 1996, 2005; Tsurutani et al., 2007, 2008; Cliver & Dietrich, 2013; Hayakawa et al., 2017c; Lakhina & Tsurutani, 2017).

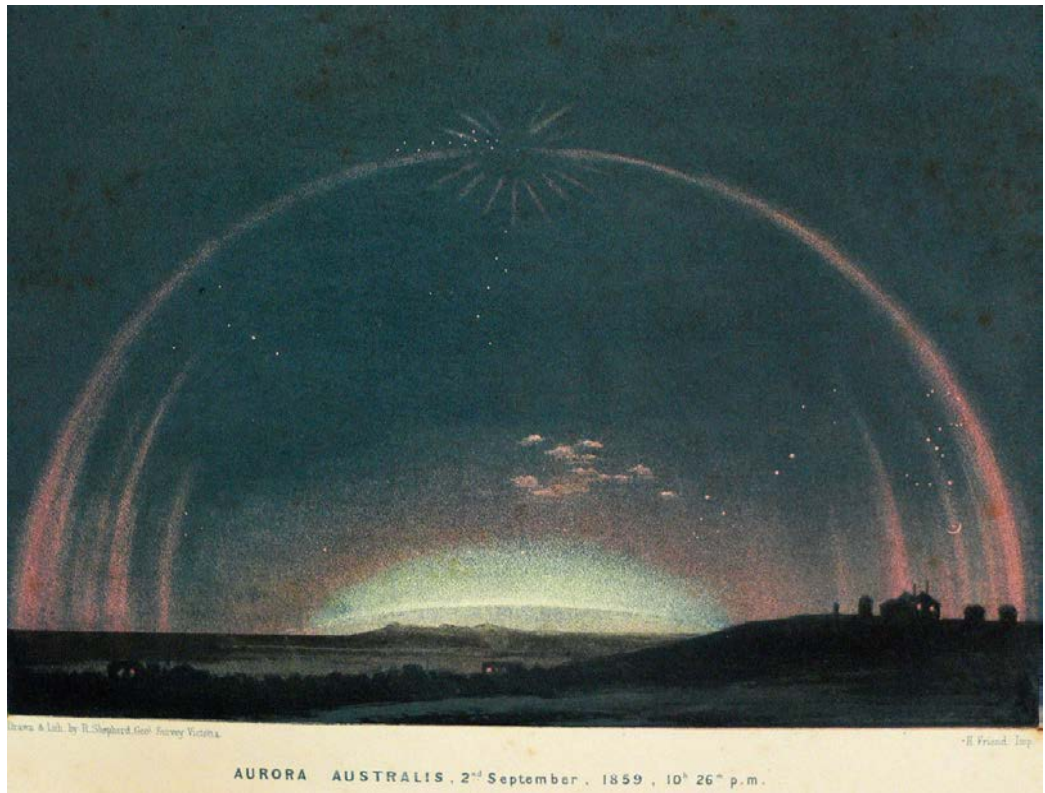

Figure 1: A drawing of auroral display with corona at Melbourne Flagstaff Observatory (S37°49′, E145°09′; -47.3° MLAT) at 22:26 on September 2, 1859, reproduced from Neumeyer (1864). Neumeyer (1864, p.242) describes this auroral observation as "At 10.26 p.m. the light of stars of the third and fourth magnitude very much enfeebled. Beautiful rays through "Pisces". During the last 10 or 15 minutes a beautiful red arc of light, extending from E. to W., and passing through the crown, had become almost stationary. It followed the astronomical equator to a height of 70° where it deviated towards south". This drawing is reproduced in Cliver and Keer (2012) as well.

The auroral records during this stormy interval have been surveyed and re-discovered after Kimball (1960) also. So far, the records such as those in US ship logs (Green et al., 2006; Green and Boardsen, 2006), American newspapers (Odenwald, 2007), Australian reports (Humble, 2006), Spanish newspapers (Farrona et al., 2011), historical documents in East Asia (Willis et al., 2007; Hayakawa et al., 2016a), and Mexican newspapers (Gonzalez-Esparza and Cuevas-Cardona, 2018) have been surveyed. These re-discovered records have provided further insights upon the auroral displays during this stormy interval.

These magnetic storms caused one of the earliest space weather disasters or space storms (see also Daglis (2003) for the terminology), such as a disturbance in the telegraph system (e.g. Loomis, 1861b, 1865). Boteler (2006) and Muller (2014) summarized glitches of telegraph transmissions, and showed that the telegraph operations were disrupted in North America and Europe on August 28/29 and September 1/2-2/3. Due to the increasing dependence upon the electricity and electronics, our society becomes increasingly vulnerable to the space weather disasters or space storms (Daglis, 2000, 2004; Baker et al., 2008). Had it occurred in the present time, the consequences are thought to be disastrous for a modern civilization that depends on electronic devices, while this detail is still controversial (Baker et al., 2008; Hapgood, 2011, 2012; Cannon et al., 2013; Oughton et al., 2016; Riley et al., 2018). Therefore, research on such extreme magnetic storms is important in geophysics and solar physics, as well as in various other scientific fields (e.g., Schwenn, 2006). In this context, it is also discussed how frequently such extreme magnetic storms occurred (e.g. Willis et al., 1997; Love, 2012; Riley, 2012; Schrijver et al., 2012; Yermolaev et al., 2013; Usoskin & Kovaltsov, 2013; Shibata et al., 2013; Cliver & Dietrich, 2013; Maehara et al., 2015; Curto et al., 2016; Riley & Love, 2017), whereas their methodologies and predictions vary from each other.

It is known that the auroral oval moves equatorward, and the aurorae dominated by red color appear in middle and low-latitude areas during magnetic storms (Tinsley et al., 1986; Shiokawa et al., 2005). The magnetic latitude of the equatorward boundary of the auroral oval is correlated with the disturbance storm time (Dst) index (Yokoyama et al., 1998). The Dst index is used as a measure of the magnetic disturbance. Thus, the equatorward boundary of the auroral emission region may be used as a proxy measure for a magnetic storm when geomagnetic field data are unavailable. Note that further re-discovery of auroral records in lower magnetic latitude can always update this estimation. To minimize the uncertainty, we prefer to determine the equatorward boundary of the auroral *emission region*, rather than determining the equatorward boundary of auroral *visibility*.

During the interval between August 28 to September 4, 1859, the equatorward boundary of auroral *visibility* has been thoroughly studied, but their exact values remained somehow controversial (e.g., Kimball, 1960; Tsurutani et al., 2003; Green & Boardsen, 2006; Cliver & Dietrich, 2013); On one hand, Kimball (1960) states that red glows were visible down to 22~23° MLAT on September 1/2 and adopted by Tsurutani et al. (2003). Green & Boardsen (2006) concluded that the aurorae were visible as low as ~18° MLAT on September 2/3, and ~25° MLAT on August 28/29, 1859.

As for the equatorward boundary of auroral *emission region*, Kimball (1960, see Figure 6) considered that "overhead aurorae" were coming down to 34~35° MLAT, while "southern extent of visibility" down to 22~23° MLAT in the Eastern United States on 1/2 September, 1859. Considering that the equatorward boundary of auroral *oval* is a better measure than the equatorward boundary of auroral *visibility* to scale magnetic storms (Yokoyama et al., 1998), we believe that reevaluating the equatorward boundary of auroral *emission region* is important to scale the magnetic storms in the stormy interval around the Carrington storm more precisely.

It should be also noted that stable auroral red arcs (SAR arcs) are frequently visible as reddish glows a few degrees equatorward of the auroral oval (Rees and Roble, 1975). The SAR arcs are typically observed during the storm recovery phases (Shiokawa et al., 2005), and are thought to coincide with the interaction region between the plasmapause and the inner edge of the ion plasma sheet (or the ring current) (Cornwall, 1970, 1971; Kozyra et al., 1997). Therefore, the equatorward boundary of the SAR arc may provide a rough estimate of the inner edge of the ion plasma sheet (or the ring current). Tsurutani et al. (2003) assumed that the purely red emission corresponds to SAR arcs, and estimated the inner edge of the ring current. With the location of the ring current, Tsurutani et al. (2003) estimated the magnetospheric electric field and used this value to obtain the Dst value for the Carrington storm. Hereinafter, we use the term "auroral emission region" instead of using "auroral oval" because we cannot exclude the possibility of the SAR arcs.

In the reevaluation of the equatorward boundary of the auroral emission region, there are some difficulties as follows. Firstly, majority of previous studies have only considered the magnetic latitude of observational sites and have not considered the auroral elevation angle therein. We need to consider the elevation angle of auroral display in eyewitness reports in low magnetic latitude, to reconstruct the equatorward boundary of auroral emission region. Secondly, the exact location of the most equatorward observational sites of Green & Boardsen (2006) is not very clear. Green & Boardsen (2006) seem to rely on their equatorwardmost observations written in the ship deck log in NARA, as shown in their Table 1 and Figures 1 and 2. Table 1 dates the observations from Panama on August 29, 1859, while Figures 1 and 2 place the observations from Panama on "September 2-3." On the contrary, Green et al. (2006) show that all of these records are dated on August 28-29, 1859. Thirdly, the difficulty also arises from the fact that the auroral emission extends from ~100 km to ~400 km along a magnetic field line that is highly inclined at low magnetic latitudes. In this paper, we re-evaluate the equatorward boundary of the auroral emission region during this stormy interval from August 28 and September 4, 1859 on the basis of eyewitness reports of auroral displays with their elevation angle, color, and brightness, according to the historical documents in time.

## 2 Materials and Methods

In order to visit the magnetic storms which occurred between August 28 and September 4, 1859, we examine contemporary source documents for eyewitness auroral reports (see the references in Appendix 1) from low-latitude areas (< 35° MLAT). The first document is the contemporary auroral reports compiled by Loomis, which primarily covers the western hemisphere and were once catalogued by Kimball (1960). After the intense auroral display in 1859, Loomis called for the eyewitness reports upon the readers of American Journal for Science to collect worldwide auroral reports from the western hemisphere. The second document contains the auroral reports from U.S. Navy ship logs. This record group was introduced by Green & Boardsen (2006) and Green et al. (2006) and formed the backbone of their discussion. The third is the historical documents in East Asia that were introduced by Hayakawa et al. (2016a), with one more record that was found after its publication (HJ5). As contemporary East Asian residents did not understand the physical nature of auroral displays, these records are not found in scientific accounts, but rather in the diaries or chronicles from these countries (see, Hayakawa et al., 2016). The fourth is the reports in Mexican newspapers, recently re-discovered by Gonzalez-Esparza and Cuevas-Cardona (2018).

We analyze this extreme auroral display in the Carrington storm based on its elevation angle, color, and brightness, as well as based on eyewitness reports with dates from low-latitude areas (< 35° MLAT) in these contemporary source documents. We first compute the magnetic latitude of the observation sites. We define the magnetic latitude as the angular distance from the dipole axis. The dipole axis is determined by using the geomagnetic field model GUFM1 (Jackson et al., 2000). When the geographical coordinates of observational site are not given in original documents, we estimate the location of observational sites as the old town/city areas in given sites, unless otherwise endorsed, considering the development of town/city area in mid 19th century (e.g. Ezcurra and Mazari-Hiriart, 1996). In this process, we revised some of the geographical coordinates of Kimball (1960) and Gonzalez-Esparza and Cuevas-Cardona (2018). Especially, we found the observational sites reported by Gonzalez-Esparza and Cuevas-Cardona (2018) are somewhat located westward from the known old town/city area by 6~80km. For example, Gonzalez-Esparza and Cuevas-Cardona (2018) located Mexico City as the geographical coordinate as "19.39 (latitude) and −99.28 (longitude)" corresponding to the location of current San Fernando District, while the Mexico City had not expanded enough to cover this district in mid 19th century, as seen in Figure 2 of Ezcurra and Mazari-Hiriart (1996). Therefore, we revised the geographic coordinates of observational sites so as to correspond to the old town, unless otherwise endorsed. Note that we have not included the report of Michoacán and San Luis Potosí in Gonzalez-Esparza and Cuevas-Cardona (2018), as they are without the exact date in their source document. Likewise, we have not included the report of Montería in Columbia in 1859 (Moreno Cárdenas et al., 2016), as this report was originally dated as "Marzo (March)" without exact date in 1859 and its dating is not clear (Exbrayat, 1971, p.151).

We then extract information on elevation angle, color, and brightness from the original eyewitness reports. We use the information on elevation angle of auroral display to determine the equatorward auroral extension by geometric calculation. We then analyze the distribution of auroral color to determine what kind of elements were influenced by incident electron particles, and finally apply the simulation code by Ebihara et al. (2017) to reconstruct the auroral brightness during this magnetic storm. Finally, we compare their duration with contemporary magnetic observations taken from the Colaba Observatory in India (Moos, 1910a, 1910b; Tsurutani et al., 2003; Kumar et al., 2015) and magnetic observatories from the contemporary Russian Empire (e.g., Nevanlinna, 2004, 2006, 2008).

## 3. Equatorward Extensions of the Auroral Emission Region and Visibility between August 28 and September 4, 1859

### 3.1. Estimation of Equatorward Boundary of Auroral Emission Region

After assembling the eyewitness reports from Loomis's collection, U.S. Navy ship logs, and East Asian historical documents, we extract observation sites with the lowest magnetic latitude of observational sites (< 35° MLAT). To estimate the equatorward boundary of the auroral emission region, we need to know the information about the elevation angles in these reports. The "overhead aurora" as presented in Figure 6 of Kimball (1960) can be regarded as the elevation angle of 90°. We should note that the instantaneous distribution of the auroral displays depends on time and magnetic local times. It is our intention to find the equatormost extension of the auroral emission region, not to find the spatiotemporal evolution of the auroral emission region.

During this stormy interval, 9 records are found to contain information about the elevation angle of the aurora as listed in Table 1. Assuming the height of the upper border of the aurora, we estimate the equatorward boundary of the auroral emission region on the basis of the geometry of the dipole magnetic field line as shown in Figure 2 (See also Hayakawa et al. (2018c)). We also assume that (1) the aurora is very thin in the latitudinal direction and is extended along a dipole magnetic field line, and that (2) atmospheric refraction is negligible. The magnetic latitude of the aurora $\lambda$ at height $h$ can be computed using the following equation for a given elevation angle $\beta$ and magnetic latitude of the observation site $\lambda_0$:

$$(a+h)\cos(\lambda-\lambda_0)=a+(a+h)\sin(\lambda-\lambda_0)\tan \beta, \qquad (1)$$

where $a$ is Earth's radius. With the dipole magnetic field, we can compute the magnetic latitude of the magnetic footprint of the aurora $\Lambda$ as

$$\Lambda= \cos^{-1}(\cos \lambda(a/(a+h))^{1/2}). \qquad (2)$$

$\Lambda$ is referred to as an invariant latitude (ILAT), which is associated with the $L$-value ($\equiv 1/\cos^2\Lambda$). Hereafter, we evaluate the equatorward boundary of the auroral emission region in terms of ILAT.

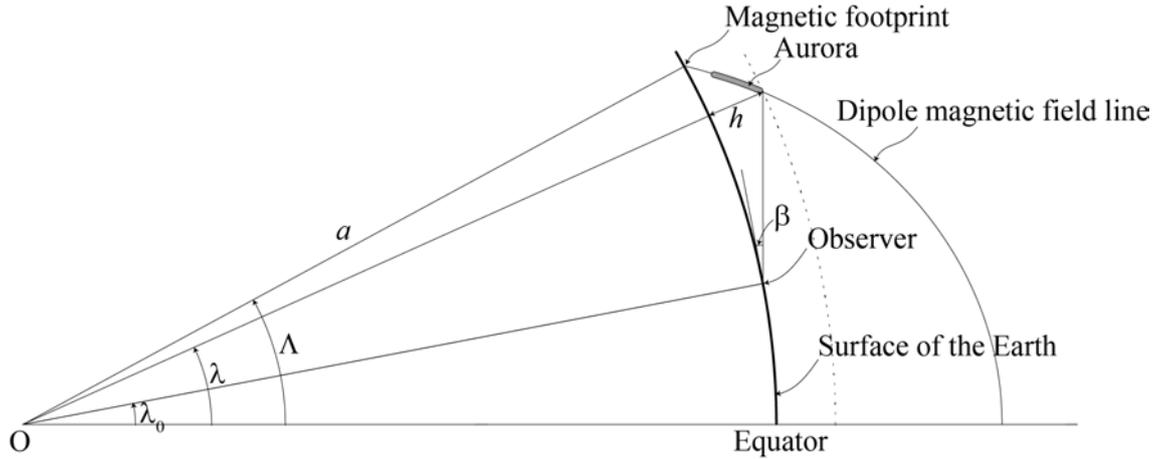

Figure 2: Relationship between the elevation angle of the auroral display $\beta$, and the invariant latitude of the aurora $\Lambda$ in dipole geometry.

In reality, the aurora has finite thickness. When the observer is located equatorward of the aurora ($\Lambda > \lambda_0$), the top border of the aurora provides an estimate of the equatorward boundary of the aurora regardless of the latitudinal thickness of the aurora.

The altitude of the upper border of the aurora is problematic because the volume emission rate of the aurora gradually decreases with altitude and there is no clear border. According to Monte Carlo simulation, the peak altitude of the volume emission rate at 630.0 nm [OI] is ~350 km and ~270 km for the precipitating electrons with monochromatic energy of 100 eV and 500 eV, respectively (Onda and Itikawa, 1995). Of course, the simulation result depends on energy and pitch angle distributions of precipitating electrons as well as temperature of electrons, ions and neutrals, and atmospheric constitution (Solomon et al., 1988). The precise altitude of the volume emission rate is not our focus because the description of the aurora has no precise information about the altitude distribution of the brightness. In reality, the distribution function of the precipitating electrons is not monochromatic, and the altitude profile of the volume emission rate depends on the distribution function of the precipitating electrons. Ebihara et al. (2017) surveyed the distribution function of the precipitating electrons measured by the DMSP satellites for severe magnetic storms, and identified two components of the distribution function. One component peaks at ~70 eV and the other one peaks at ~3 keV. According to the two-stream electron transport code used by Ebihara et al. (2017), the volume emission rate at 630.0 nm peaks at ~270 km for the electron distribution function measured in the severe storms. The altitude at one tenth of the maximum volume emission rate occurs at ~410 km altitude.

## 1859 August 28/29

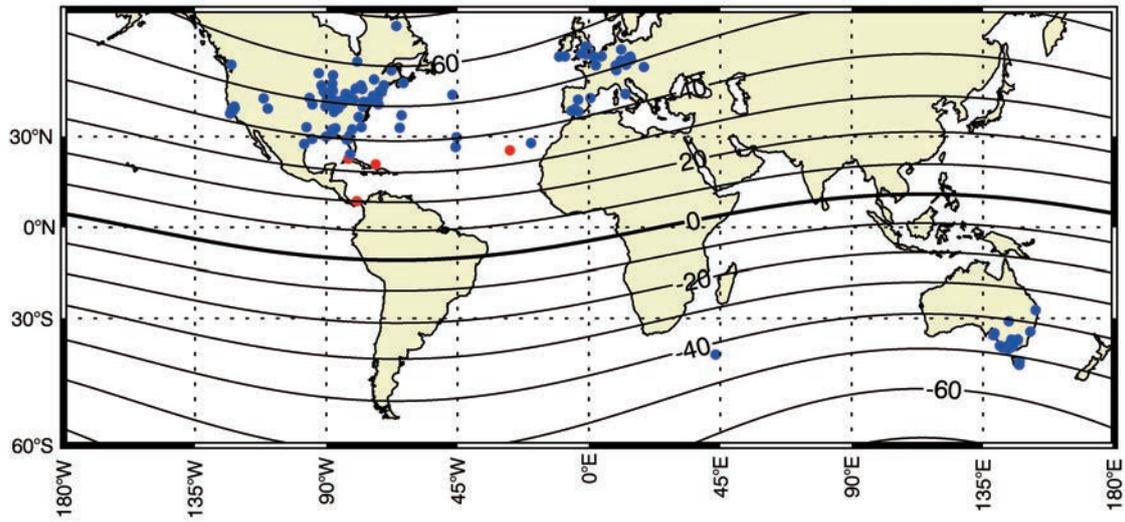

## 1859 September 1/2-2/3

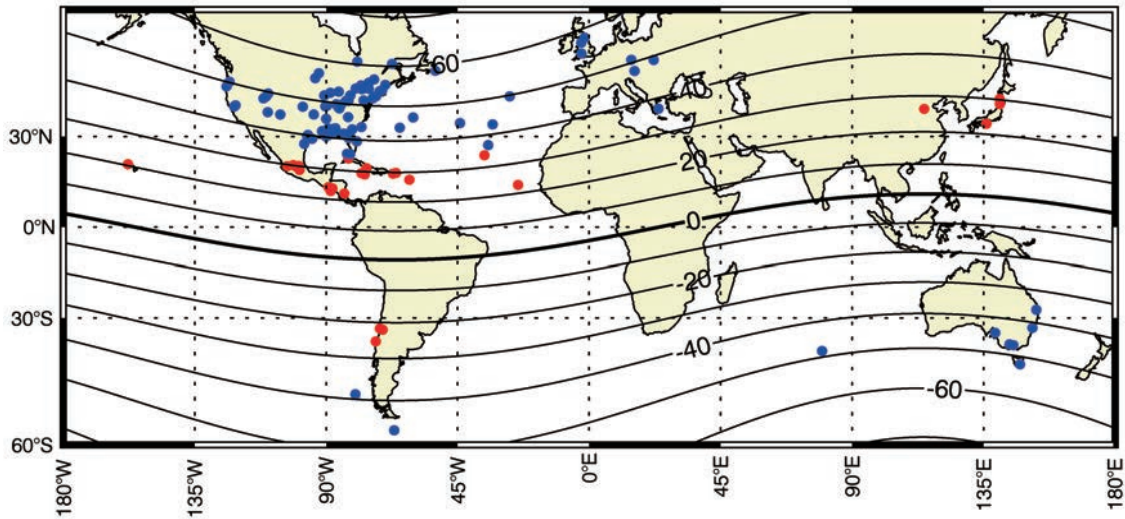

Figure 3: Observational sites of auroral displays during August 28/29 (a: top) and September 1/2 (b: bottom), 1859. Only the sites within ±60 ° are shown. The sites with their magnetic latitude being lower than 35° are shown in red (see, Table 1). The sites with their magnetic latitude being higher than 35° are shown in blue. The latter observational sites are based on reports by Loomis (L1-L8), WAMG (Heis, 1859, 1860), Neumeyer (1864), and references and data reductions in Humble (2006), Green and Boardsen (2006), and Farrona et al. (2011). The contour indicates the magnetic latitudes in 1859 calculated on the basis of the GUFM1 magnetic field model.

### 3.2. Auroral Display During August 28/29

Table 1 and Figure 3a summarize these observation sites at the magnetic latitude being less than 35° on August 28/29. During these days, the most equatorward site of *auroral visibility* is 20.2° MLAT (Panama; Saranac, 1859-08-29), while Green & Boardsen (2006) concluded the equatorward boundary of this *auroral visibility* as 25° MLAT. Note that Green and Boardsen (2006) dated the record at Panama on August 29 in their Table 1, while they dated it not on August 28/29 but September 2/3 in their Figures 1 and 2.

In the same time, Fritz (1873) mentioned that aurora was visible at St. George del Mina, namely current El Mina (N05°05′, W001°21′) on August 28, 1859. If this were the case, aurora would have been visible down to 9.9° MLAT according to the GUFM1 model (Jackson et al., 2000). However, the original text shows that it was Dr. Daniels who wrote "from St. George del Mina (West Coast of Africa)" who witnessed the auroral display "at 19°50′ west of Greenwich and 28° north" (WAMG, v.3, p.38 = WA2). Therefore, his observation (WA2) took place not at St. George del Mina (N05°05′, W001°21′; 9.9° MLAT), but at the sea near Cabo Verde (N28°, W19°50′; 35.5° MLAT), as also suggested by Silverman (2008). Therefore, the equatorwardmost observational site (*visibility*) on August 28/29 should be located at Panama (20.2° MLAT).

Nevertheless, it does not necessarily mean that the equatorward boundary of auroral emission region came to the zenith of Panama. The record of Saranac at Panama (20.2° MLAT) does not provide the information about elevation angle of this auroral display a little before 04:00 LT. However, we find auroral display "rising to the zenith" at Havana (34.0° MLAT) around 04:00-04:10 LT on August 29 (L1, pp.403-404 = L1-8) When we simply assume that the upper border of the aurora is 400 km altitude, we can estimate the equatorward boundary of the auroral emission region to be 36.5° ILAT (or 34.0° MLAT at 400 km altitude). In this case, the auroral display can be seen up to an elevation angle of 7° at Panama (20.2° MLAT).

### 3.3. Auroral Display During September 1/2-2/3

Table 1 and Figure 3b summarize these observation sites at the magnetic latitude being less than 35° during September 1/2-2/3. On September 1-2, the two most equatorward sites are −21.8° MLAT (Valpalaiso, Chile; L4, p. 399 = L4-15) and 22.8° MLAT (at sea; L6, p. 361 = L6-2-43 and WAMG, v.3, p.270 = WA1). We excluded the record at Honolulu because the exact date of auroral observation is not provided in the original document (L5, p. 88). We must also note that another cluster of observation sites are found in East Asia down to 23.1° MLAT (Shingu, Japan).

The equatorwardmost report having information about elevation angle comes from Sabine (23.1° MLAT). The record of Sabine is not included in Kimball (1960), but is consistent with his result in term of the latitudinal extent of auroral visibility. The report from Sabine shows that auroral display extended up to 35° in elevation angle from 00:30 LT to 01:30 LT on September 2 (Sabine, 1859-09-02). Assuming the auroral height of ~400 km and substituting $\beta$ of 35° and $\lambda_0$ of 23.1° into Eqs. (1) and (2), we estimate the equatorward boundary of the auroral emission region to be 30.8° ILAT (or 27.7° MLAT at 400 km altitude). Likewise, another naval record by Captain Kraan (WA1) shows reddish aurora was visible up to 30° in elevation at the sea (N14°28', W024°20') and lets us estimate the equatorward boundary of the auroral emission region to be 31.3° ILAT (or 28.2° MLAT at 400 km altitude) between 4:30 LT and 5:15 LT on Sept. 2. These records are consistent with the record at Porto Rico (29.8° MLAT) in which "luminous rays, red, purple and violet, extended even to the zenith" (L5, p. 88 = L5-14) almost simultaneously (at ~07 UT on September 2, 1859) and that at Havana (34.0° MLAT) "which passed the zenith towards the northeast, attaining the height of 100 degrees, accompanied with whitish rays and also with the red rays, more vivid then the general tones of the segment rising to the zenith, yet without passing it" (L1, p.405 = L1-8). Based on these reports, the equatorward boundary of auroral emission region is estimated 32.7° ILAT and 35.4° ILAT, respectively.

The recently recovered Mexican reports also support this estimation. Reports from Mexico City (MX1, 28.8° MLAT), Querétaro (MX2, 29.8° MLAT), and Zimapán (MX4, 30.1° MLAT) show that the auroral displays were reaching to the zenith, and suggest that the equatorward boundary of the auroral emission region came down to 31.8° ILAT, 32.7° ILAT, and 31.9° ILAT, respectively. The report from Zimapán is interesting as it may possibly refer to auroral corona, mentioning "a silver lily in the shape of an arc of a great circle" from the region where "glowing rays extended downwards as if to meet a red light that shone up from the northern horizon" (MX4, see also Gonzalez-Esparza and Cuevas-Cardona, 2018).

If the historical description in Honolulu (20.5° MLAT), in which the aurora extended up to 35° from the horizon, mentions the aurora found on September 1-2, the auroral oval during this interval would be extended to 28.5° ILAT (or 25.1° MLAT at 400 km altitude). In short, these remote and independent observation sites of low-latitude aurorae suggest the following:

1.  The equatorwardmost magnetic latitudes (MLATs) of the auroral *visibility* are ~20.2° during the August 28/29 storm, and −21.8° and 22.8° during the September 1-2 storm.

2.  The equatorwardmost invariant latitudes (ILATs) of the auroral *emission region* are 36.5° during the August 28/29 storm, and 30.8° during the September 1/2 storm.

These latitudes are lower than those estimated by Kimball (1960). Moreover, the equatorward boundary of the auroral emission region during the September 1/2 storm, obtained here, lets us compare this event with another rivaling extreme event on February 4, 1872 (Chapman and Bartels, 1940; Chapman, 1957; Cliver and Svalgaard, 2004; Tsurutani et al., 2005; Silverman, 2008). During this storm, aurora was observed at Bombay (10.0° MLAT) as Chapman and Bartels (1940) noted (Tsurutani et al., 2005; Silverman, 2008). Recent surveys in the East Asian sector showed that the aurora was observed at the zenith of Shanghai (19.9° MLAT), reconstructed its equatorward boundary of auroral emission region as 24.2° ILAT, and estimated that the auroral display would have been indeed visible at Bombay within the elevation angle of 10°-15° (Hayakawa et al., 2018b). Further studies would be in need to compare these extreme space weather events suggested by Chapman (1957).

## 4 Color of the auroral display

As shown in Table 1, the auroral display during the stormy interval basically shows red color, but some of them show other colors. For example, on August 28-29, whitish auroral displays were observed at Havana as well (34.0° MLAT) in the northern hemisphere (L1, pp. 403-404). On September 1/2, the observer at Porto Rico (29.8° MLAT) noticed "luminous rays, red, purple and violet, extended even to the zenith" (L5, p. 88), and another observer at Guadeloupe (27.5° MLAT) "noticed two rays of whitish light which rose parallel to each other, passing a little to the left of the pole star" (L3, p. 265). In the southern hemisphere, an observer at Santiago noted "brilliantly illuminated by a light, composed of blue, red, and yellow colors, which remained visible for about three hours" at 02:00 LT on September 2, 1859 (L4, p.399).

These colorful auroral displays including whitish one may suggest the existence of greenish aurorae (557.7 nm [OI]) and/or bluish aurorae (427.8 nm [N$^+$]) caused by precipitation of electrons with energy above ~1 keV, in addition to above-mentioned red aurorae. The mixture of these auroral displays may explain the whitish and yellowish aurorae as well. A similar description is found in the much earlier historical document for auroral displays in 771/772 and 773, with rays or scepters in colors of "blood-red, green, and saffron-colored" observed at Amida (45° MLAT) according to the Zūqnīn Chronicle (MS Vat. Sir. 162, f.150v, f.155v) (Hayakawa et al., 2017b). The cause of the precipitation of the electrons remains an open question.

The appearance of the rays with whitish, red, purple and violet colors may result from a fold of a sheet-like structure of aurora (Oguti, 1975). When the sheet-like structure of aurora is folded, the line-of-sight integral of the light is increased, resulting in a localized enhancement of brightness at all wavelengths. If this was the case, the formation of the sheet-like aurora in the low-latitude aurora dominated by red color would be a problem because such a sheet-like red-dominated aurora is unusual. If the rays were caused by a localized enhancement of electron precipitation, magnetospheric processes would be a problem because localized precipitation of low-energy electrons is unusual at low-latitude. Anyway, the records in Table 1 raise a new problem regarding low-latitude aurora. The deep inner magnetosphere ($L<1.5$) may be much more complicated than we believe.

Most of the records in Table 1 indicate that the aurora is dominated by red color (most likely 630.0 nm [OI]). With optical measurements with a bandpass filter, Miyaoka et al. (1990) and Shiokawa et al. (2005) have shown that there are two types of red display observed in Japan. One is the red-dominant display with emission at 557.7 nm (e.g., 21 October 1989 and 29-30 October 2003). The energy source for the red-dominant display is precipitation of low energy electrons (<~100 eV) (Banks et al., 1974). The precipitating electrons excite the atomic oxygen to the $O(^1D)$ state, and the transition OI ($^3P$-$^1D$) results in emission at 630.0 nm (Rees and Roble, 1975). The transition OI ($^1D$-$^1S$) gives rise to emission at 557.7 nm. The excitation energies of the $OI(^1S)$ and $O(^1D)$ states are 4.19 and 1.97 eV, respectively (Rees and Roble, 1975). The emission at 630.0 nm dominates that at 557.7 nm because the probability of the $O(^1D)$ state is about 10 times higher than that of the $OI(^1S)$ state (Rees, 1989). If this was the case, electrons would originate from adiabatically accelerated plasmaspheric populations (Ebihara et al., 2017). The other type is the red display without discernible emission at 557.7 nm (e.g., 7 April 2000). This can be regarded as a SAR arc (Roach and Roach, 1963). The energy source of the SAR arc is heated electrons (~3000 K) associated with heat flows from high altitude, or very low energy particle flux (Cole, 1965; Cornwall et al., 1970, 1971; Kozyra et al., 1997). The thermal electron flux decreases with energy, which also gives rise to the dominance of 630.0 nm (Kozyra et al., 1997). Very bright SAR arcs with intensity up to 13 kilo Rayleighs were observed when a large magnetic storm occurs (Baumgardner et al., 2007). There are at least 3 processes for the energy conversion from the magnetospheric ions to the ionospheric electrons (Kozyra et al., 1997), including Coulomb collision (Cole, 1965), wave-particle interaction (Cornwall et al., 1971), and kinetic Alfvén waves (Hasegawa and Mima, 1978). If this was the case, the red-dominant display corresponds to the footprint of the interaction region between the storm-time ring current and the plasmasphere. Mendillo et al. (2016) show an example that the red aurora (~200 km altitude) and the SAR arc (~400 km altitude) coexist on the same field line. It should be noted that Tsurutani et al. (2003) used the equatorward visibility (~23° MLAT) of these "red glows" in Kimball (1960) to scale this magnetic storm in comparison with the magnetic observation at Colaba.

We consider that it is not straightforward to distinguish between the aurora and the SAR arcs from the existing records, due to the lack of objective records by scientific equipment. Considering formless features and red-dominated color of the SAR arcs with relatively longer duration (Cornwall et al., 1970, 1971; Kozyra et al., 1997), one may consider the red-dominated displays without other colors, or motions and with longer duration at low magnetic latitude to be SAR arcs (K. Shiokawa, private communication). Forms are not good criteria to distinguish them because structured SAR arcs are observed (Mendillo et al., 2016). As listed in Table 1, some records fit these criteria at, for example, La Union (L3-31, 23.8° MLAT), Kingston (L3-29, 29.1° MLAT), and Montego Bay (L3-29, 29.5° MLAT). These are likely to be SAR arcs because the reddish aurora had been observed without motion for ~4-5 hours. With information about elevation angle at La Union (L3-31, 23.8° MLAT), we estimated that the equatorward boundary of the aurora extended to 32.2° ILAT. The most equatorward boundary of the aurora with red color only was located down to 30.8° ILAT (at Sabine, RG24-2). They are also presumably considered as SAR arcs, unless otherwise some typical motions or structures are mentioned.

In the same time, there are some auroral reports unlike SAR arcs even down to lowest magnetic latitude, possibly related with auroras by broadband electrons (e.g. Shiokawa et al., 1997, 1999). The auroras with multiple colors are reported down to 29.8° MLAT with "red, purple and violet" colors (Porto Rico) and −22.1° MLAT with "blue, red, and yellow colors" (Santiago, L4-15). The equatorward boundary of the auroral emission region with multiple colors is estimated to be 32.7° ILAT (Porto Rico, L5-14). They are not likely SAR arcs as they have non-reddish components. Likewise, a report at 20.5° MLAT (Honolulu) describes "Broad fiery streaks shot up into and played among the heavens" in it, while their dating is uncertain. This is likely ray structure of type A aurora (Chamberlain, 1961), rather than SAR arcs. If we can date this record as on September 1 as in Kimball (1960), the equatorward boundary of auroral *oval* would be calculated even down to 28.5° ILAT, considering its elevation angle ~35°. These reports show not only SAR arcs but also usual auroras were distributed even down to the most equatorward in terms of their visibility.

The equatorward boundary of the auroral *oval* may provide the upper limit of the inner edge of the electron plasma sheet (e.g., Vasyliunas, 1970). Horwitz et al. (1982) show two examples indicating that the equatorward boundary of the auroral oval coincides with the inner boundary of the plasma sheet and the plasmapause. On the other hand, the equatorward boundary of the SAR arcs may reflect the interaction region between the plasmasphere and the inner boundary of the ion plasma sheet (or the storm-time ring current) (Cornwall et al., 1970, 1971; Kozyra et al., 1997). The earthward transport of the electron plasma sheet and the ion plasma sheet are most likely caused by the enhancement of the large-scale convection electric field. The magnetospheric convection is enhanced when the southward component of the interplanetary magnetic field (IMF) arrives at the Earth (e.g., Kokubun, 1972).

The strong convection electric field transport fresh electrons originating in the nightside plasma sheet toward the Earth by the E×B drift. When the electrons experience pitch angle scattering, some of them are scattered into the loss cone, resulting in the diffuse aurora (e.g., Lui et al., 1977). The equatorward boundary of the (diffuse) aurora is reasonably supposed to provide an upper limit of the earthward boundary of the electron plasma sheet. The earthward boundary of the electron plasma sheet is determined by the strength of the convection electric field, and is located at, or outward of the plasmapause (Ejiri et al., 1980).

Fresh ions originating in the nightside plasma sheet are also transported earthward as previously mentioned by Tsurutani et al. (2003) for the Carrington storm. Because of energy-dependent drift velocity, the inner edge of the ion plasma sheet depends on the particle perpendicular kinetic energies. Observations have shown that ions at particular energies can penetrate into the plasmapause (Smith and Hoffman, 1974). The energy-dependent penetration of the ions is called a nose structure (Smith and Hoffman, 1974), and is theoretically explained by Ejiri et al. (1980). When the ions at particular energies interact with the plasmaspheric cold plasmas, temperature of the plasmaspheric cold electrons increase. Consequently, the electron temperature increases in the topside ionosphere, causing the SAR arc emissions (Cornwall et al., 1970, 1971; Kozyra et al., 1997). We note that the inner edge of the ion plasma sheet may have some ambiguity of a few degrees in magnetic latitude because the equatorward boundaries of the ion plasma sheet depend on energy (Ejiri et al., 1980). The equatorward boundary of the SAR arc may provide a rough estimate of the inner edge of the plasma sheet, while may have some ambiguity.

In general, the SAR arcs cannot be explicitly distinguished from the usual aurora caused by energetic electron precipitation without spectroscopic instruments. Miyaoka et al. (1990) present a photograph of the low latitude aurora dominated by red. The picture shown by Miyaoka et al. (1990) looks purely red aurora. However, according to data from a scanning photometer, the red-dominated aurora contains emission at 557.7 nm (green), which means that the red-dominated aurora is most likely an aurora, not a SAR arc. Thus, we cannot conclude, at the present stage, that all the purely red emission found in the historical records corresponds to the SAR arcs.

## 5 Abnormal auroral brightness during the Carrington storm

Intense electron precipitation can cause much brighter aurorae than normally expected at low latitudes. The Baltimore American and Commercial Advertiser on September 3, 1859 (p. 2, col. 2) describe the magnificent auroral display "on late Thursday night" (September 1, 1859) and conclude the auroral "light was greater than that of the moon at its full." Other similar descriptions are found in the low-latitude areas (< 35° MLAT), as summarized in Table 1. At Guadeloupe (27.5° MLAT), "its ruddy light was noticeable in the interior of the houses" (L3, p. 265). At La Union (23.8° MLAT), the auroral display was described "light was equal to that of day-break, but was not sufficient to eclipse the light of the stars" (L3, p. 265). At Concepcion (−25.5° MLAT), the auroral display "threw out some flame or vapor, and spread a light like that of the moon" (L4, pp. 398-399). Other records generally compare this auroral display with a conflagration or colossal fire, especially in East Asia (see, Hayakawa et al., 2016a).

According to the above-cited descriptions, we consider the auroral brightness to be Class IV International Brightness Coefficient (IBC), where the total illumination on the ground equals to that of full moon (Chamberlain, 1961). IBC Class IV is suggested to correspond to a brightness of approximately 1000 kilo Rayleigh (kR) for the "green aurora" at 557.7 nm (Hunten et al., 1956). As far as we know, such bright aurorae have probably not been recorded by modern scientific instruments at low latitudes (Hikosaka, 1958; Shiokawa et al., 2005). Similar bright aurorae have been recorded in East Asia but in 1770 before magnetic observations, which are described as "as bright as a night with full moon" at Nagoya, Japan (N35°11′, E136°54′, 25.2° MLAT) (Ebihara et al., 2017). Unusually intense electron precipitation, about an order of magnitude larger than that observed in the March 14, 1989 storm, is expected to cause the bright aurorae corresponding to Class IV (Ebihara et al., 2017).

SAR arcs are observed to be as bright as 13 kR during the large magnetic storm (Baumgardner et al., 2007). If the purely red emission corresponds to the SAR arc, the brightness will be problematic. Extremely dense ion plasma sheet (or the ring current), and/or extremely dense plasmaspheric electron population are expected to occur in 1859. This problem remains to be solved for future studies.

## 6 Geomagnetic Disturbances

### 6.1. General Overview

Figures 4 shows the duration of auroral observations together with records of observations at magnetic observatories in Helsinki (HEL: N60°10′, E24°57′), St. Petersburg (STP: N59°56′, E30°18′), Ekaterinburg (EKA: N56°49′, E60°35′), Barnaul (BAR: N53°20′, E83°57′), Nertchinsk (NER: N51°19′, E119°36′) (Nevanlinna, 2006, 2008), and Colaba in India (Tsurutani et al., 2003; Kumar et al., 2015) for August 28-29, 1859 and September 2, 1859, respectively. We converted the observation time of each record from local time (LT) to universal time (UT). Unfortunately, the magnetic field at Colaba is missing during the interval from 11 UT on 28 August to 12 UT on 29 August because the Colaba magnetic observatory did not perform observations on Sundays and holidays during the period 1847-1872 (Moos, 1910a, p.105). At 12 UT on 29 August, the deviation of the magnetic field ($\Delta H$) is −484 nT, suggesting that a magnetic storm could have commenced during this missing interval. It is reasonable to consider that this stormy interval from August 28 to September 4, 1859 is composed of, at least, two magnetic storms probably caused by ICMEs from the same active region. A similar stormy interval was observed in October 2003 during which multiple ICMEs were launched from the same flaring active region (NOAA 10486) (Yashiro et al., 2004), causing successive large magnetic storms (Shiota and Kataoka, 2016). In November 2004, multiple ICMEs were also launched from the same active region (NOAA 10696), causing successive large magnetic storms (Echer et al., 2010). One single ICME is unlikely to cause successive magnetic storms with interval of 3-4 days. One single ICME may cause two-step development of magnetic storms (Kamide et al., 1998; Daglis, 2004) when southward component of IMF is embedded both in the sheath and the magnetic cloud in the ICME (Tsurutani et al., 1988).

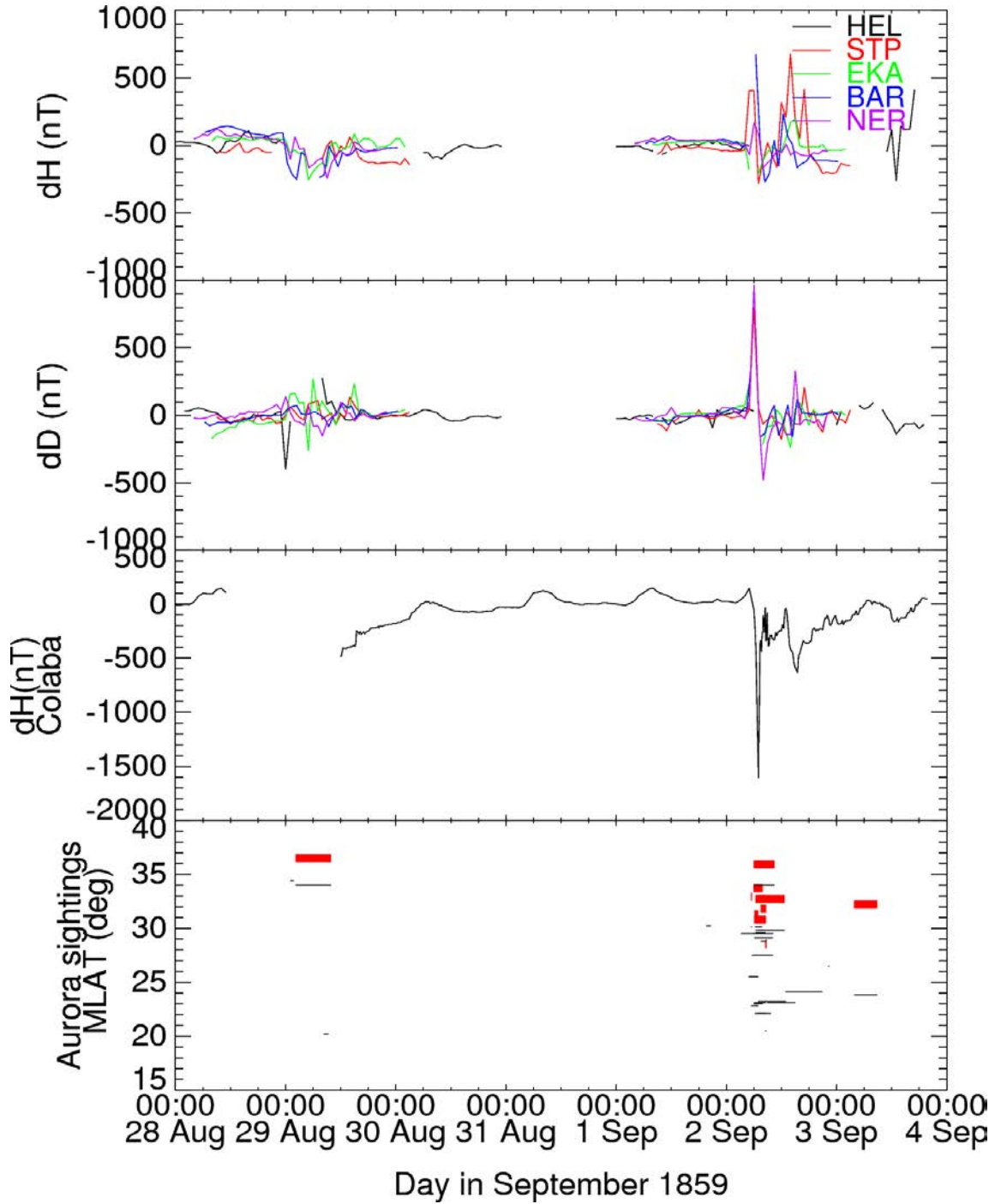

Figure 4: From top to bottom: magnetic disturbance in the H-component (magnetic north-south component) at Helsinki (HEL), St. Petersburg (STP), Ekaterinburg (EKA), Barnaul (BAR), and Nertchinsk (NER); magnetic component in the D-component (magnetic east-west component), magnetic disturbance in the horizontal component at Colaba, and time and magnetic latitude when aurora was seen from August 28 to September 4, 1859 (UT). The horizontal (red) thick line indicates the possible equatorward boundary of the auroral emission region in ILAT estimated from the auroral elevation angle. The horizontal (black) thin line indicates the MLAT at which aurora was visible. The report of Honolulu is plotted as an assumption of its being observed on September 1/2.

If multiple ICMEs launched from the same active region, the former one could be decelerated by momentum transfer or aerodynamic drag as it propagated into interplanetary space due to the "snow plough" effect (Tappin, 2006; Takahashi and Shibata, 2017). If the solar wind density is low on the trailing side of it, the latter ICME, leaving the Sun several days later, could proceed in interplanetary space without substantial deceleration, as the former one had cleared out the interplanetary mass in advance (Tsurutani & Lakhina, 2014; Shiota and Kataoka, 2016). Because of the lower "snow plough" effect, the latter one could hit the Earth's magnetosphere without substantial deceleration, becoming more geo-effective and resulting in the extreme storm on September 1/2, known as the Carrington storm. If this is the case, the former one could play an important role in causing the Carrington storm.

Carrington's sunspot observations seem to support this scenario. As reproduced in Figure 5, a large active region was already in the eastern hemisphere of the Sun on August 28 and came to the central meridian on September 1. Carrington noticed this active region on August 25 (C3, v.2, f.312a) and monitored this active region (C1, v.2, ff.63a-64a). During this observation, Carrington witnessed the white light flare on September 1 as published in Carrington (1859). In his original logbook, he highlighted the region with white light flares in reddish color (C1, v.2, f.64a). Cliver (2006) reviewed the flare on September 1 in more detail.

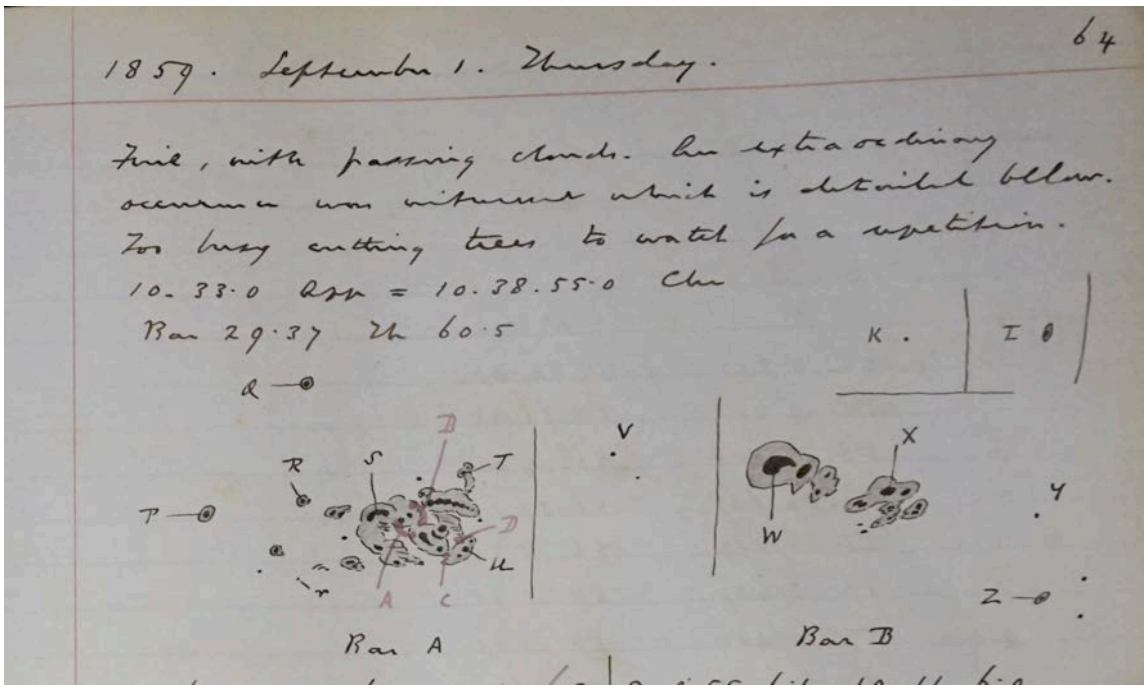

Figure 5: Carrington's sunspot drawings on August 28 and September 1, shown in projected images (see, Plate 1 of Carrington (1863) and Figure 2 of Cliver and Keer (2012)). The whole disk drawings on August 28 and September 1 are shown above (C3, v.2, ff.312a-313a). The relevant parts of his logbook on August 28 and September 1 are shown below (C1, v.2, ff.63a-64a). These manuscripts are currently preserved in the archive of the Royal Astronomical Society as shown in Appendix 1.5 (courtesy: the Royal Astronomical Society).

Typically, main phases of magnetic storms last at most a few hours in their duration in combination of sheaths and magnetic clouds (Tsurutani et al., 1988; Echer et al., 2008a). Exceptionally the Hydro Quebec storm in 1989 was reported to last for up to whole day (Allen et al., 1989). However, recent analyses clarified that multiple sheaths and magnetic clouds were combined to cause this "storm" with apparent long duration and classified this "storm" as a "compound magnetic storm" (Lakhina et al., 2012; Lakhina & Tsurutani, 2017). Likewise, in the magnetic observation at Colaba, the main phase of the Carrington storm lasted most ~1.5 hours with a fast recovery phase within these storms. It was probably caused by the strong southward component of IMF (Tsurutani et al., 2003, 2018). Compared with the "compound (multi-step) magnetic storm" of 1989, Lakhina et al. (2012) and Lakhina and Tsurutani (2017) consider the Carrington storm a "one step" storm, most probably caused by a magnetic cloud within the ICME (Kamide et al., 1998; Daglis, 2004). Moreover, the one-step Carrington storm was one of many storms that occurred between August 28 and September 4, presumably caused by multiple ICMEs from the same active region (Cliver and Dietrich, 2013).

## 6.2. Magnetic Observations During 28/29 August 1859

In Figure 4, the H-component of the magnetic field starts to decrease at 23 UT on August 28, and recovers by 12 UT on August 29 (EKA, BAR, and NER). Thus, it is speculated that the magnetic storm of 28/29 August 1859 commenced at 23 UT on August 28. This period roughly corresponds to the period of the auroral display at Havana (34.0°MLAT) from 20:45 LT on August 28 to 04:20 LT on August 29 (from 0145 UT to 0920 UT on August 29) and two maritime observations. At 0400 – 0410 LT (1029 – 1039 UT), the aurora was "rising to the zenith" at Havana. Assuming that the top side border of the aurora was located at 400 km altitude, we estimate the equatorward boundary of the auroral *oval* to be 36.5° ILAT as indicated by the horizontal thick line in Figure 4. Panama (20.2° MLAT) observed aurora around 0300 – 0400 LT (0918 -1018 UT), as indicated by the horizontal red line in Figure 4. As discussed above, the same aurora was visible in Panama and Havana. EKA, BAR, and NER were located on the dawnside. The negative excursion of the H-component is probably caused by the Hall current flowing westward associated with the DP2 current system, i.e., the ionospheric convection. If this is the case, ionospheric convection would be enhanced during this period by the intense southward component of the interplanetary magnetic field and fast solar wind. The enhanced convection electric field could result in the earthward penetration of the magnetospheric plasma originating from the nightside magnetosphere. The earthward electron penetration could result in the equatorward displacement of the auroral oval. Earthward penetration is accompanied by adiabatic acceleration of the plasma, resulting in enhanced plasma pressure in the inner magnetosphere, particularly the ring current.

After the missing interval, the magnetic field observation at Colaba resumes at 12 UT on August 29 with ΔH of −484 nT. After that, ΔH shows a gradual increase until ~00 UT on August 31. The gradual increase probably indicates a remnant of the recovery of the storm that probably initiated at 23 UT on August 28 as speculated from the high latitude observation of the magnetic field. Because of the missing interval at Colaba on August 28 (Bombay Local Time), we cannot count the number of storms during this interval, or identify the minimum of the magnetic disturbance. It is considered multiple storms probably occurred during this interval as in the "compound storm" in 1989 (Lakhina et al., 2012; Lakhina & Tsurutani, 2017), and these storms commenced at ~23 UT on August 28, with the total duration of these storms as ~49 hours.

### 6.3. Magnetic Observations During 1/2-3/4 August, 1859

On September 1-2, 1859, large amplitude, bipolar disturbances in the D-component are pronounced at least at STP and NER, in addition to the disturbances in the H-component. The horizontal component of the magnetic field is largely depressed at Colaba at 06 – 08 UT on September 2, 1859. We propose two scenarios for the magnetic disturbances on September 1-2, 1859. The first scenario is based on the spatial variation. The downward field-aligned current (dawnside part of Region 1 current (Iijima and Potemra, 1976)) located just equatorward of the observatories caused the eastward disturbance on the dawnside (e.g., STP and NER at 05 – 07 UT). The ionospheric Hall current flowing eastward caused the northward disturbance (e.g., STP and NER). This is consistent with the magnetic observations at Rome and Greenwich indicating that there was a strong westward Hall current associated with the convection (Boteler, 2006). The amplitude of the positive excursion in the H-component at NER is smaller than at STP because STP was probably closed to the throat of the dayside convection. As the Earth rotates, the observatories moved from the dawn sector to the dusk sector. The upward field-aligned current (duskside part of Region 1 current) caused the westward disturbance on the duskside (e.g., NER). The ionospheric Hall current caused the southward disturbance (e.g., STP, BAR and NER). The second scenario is based on the spatial variation. The center of the downward field-aligned current moved rapidly equatorward of the observatories, and the downward current caused the eastward disturbance on the dawnside (e.g., STP and NER at 05 – 07 UT). Then, the upward field-aligned current moved poleward rapidly. The upward current located poleward of the observatories caused the westward disturbance (e.g., STP and NER at 07 – 09 UT).

At the midday, the center of the Region 1 field-aligned current is located at 74° ~ 77° ILAT for magnetically quiet period (Iijima and Potemra, 1976). During intense magnetic storms, the center of the Region 1 field-aligned current expands as low as 50~55° ILAT on the nightside (Fujii et al., 1992; Ebihara et al., 2005). Ngwira et al. (2014) performed a global magnetohydrodynamics (MHD) simulation for a Carrington-type event based on an assumption of extremely high densities of the ICME for this event, and showed that the center of Region 1 field-aligned current is located between 40° ILAT and 50° ILAT during the main phase of the model storm on the dayside. The MHD simulation result is consistent with the variation of the D-component magnetic field in September 2-3. As for August 28-29, such bipolar variation is not clearly seen in the D-component. This probably indicates that the center of the Region 1 current was located well poleward of the observatories. Most of the auroral displays start to appear with the same timing as the sharp magnetic disturbance.

For the interval from September 2 to 3, the auroral displays were continuously seen during the main and recovery phases of the magnetic storm. This can be explained as follows. Electrons are injected earthward due to the enhanced magnetospheric convection (probably associated with the enhanced Region 1 field-aligned current). When the injected electrons are scattered by some processes into the loss cone, the aurora becomes illuminated. The seed electrons trapped in the inner magnetosphere could remain during the recovery phase, which caused the long-lasting auroral displays at low latitudes. Substorm-associated injection may also supply hot electrons deep into the inner magnetosphere. The substorm-associated injection is well observed at geosynchronous orbit, whereas is not observed in the deep inner magnetosphere, such that $L < 1.5$, as far as we know.

The September 2-3 observations of aurora at La Union and San Salvador are unusual as they are isolated in the low-latitude auroral observations. While this may be due to possible misdating of September 1 as considered by Kimball (1960), the record explicitly writes that it started "about 10 o'clock … on the night of Sept. 2d" (L3, p. 265), namely 04h UT on September 3 and corresponds to recovery phase around this time (Nevanlinna 2006, Fig. 4).

The H-component of the magnetic field recorded at Colaba shows a rapid recovery on September 2. Li et al. (2006) postulated an extremely high solar wind dynamic pressure to enhance the magnetopause current flowing in the eastward direction. One possibility is the strong magnetopause current that increases the H-component of the magnetic field. If the rapid recovery of the H-component of the magnetic field is attributed to the rapid decay of the ring current, there will be some mechanisms (Ebihara and Ejiri, 2003), including charge exchange (Dessler and Parker, 1959), resonant interaction with ion cyclotron waves (Cornwall et al., 1970, Tsurutani et al., 2018), replacement with tenuous plasma stored in the plasma sheet (Ebihara and Ejiri, 1998), and pitch angle scattering of ions in a curved field line (Ebihara et al., 2011).

The first plausible mechanism for the rapid decay is the charge exchange. Using the neutral hydrogen (geocorona) density model of Rairden et al. (1986) and the cross section model of Janev and Smith (1993), we can estimate the lifetime for the charge exchange between $H^+$ with energy of 100 keV and H to be 8.5 hour at $L$=1.5. The lifetime for the charge exchange between $O^+$ with energy of 100 keV and H is 1.0 hours with the cross section model of Phaneuf et al. (1987). Hamilton et al. (1988) suggested that the rapid recovery of the Dst index during a large magnetic storm (minimum Dst value of −306 nT) is caused by the rapid loss of $O^+$, although Kozyra et al. (1998) raised a question about the capability of the charge exchange in the rapid loss. The second mechanism is the replacement of the ring current with the tenuous one (Ebihara and Ejiri, 2003). During the main phase, a large number of ions are transported from the plasma sheet to lower L-shells by the convection electric field. If the plasma sheet density decreases rapidly, the tenuous plasma is transported to lower L-shells so as to replace the previously transported one with the freshly transported one. The energy density in the heart of the storm-time ring current decreases rapidly, and the total amount of the particles' energy decreases rapidly, which gives rise to the rapid decay of the ring current. This idea is tested by Keika et al. (2015). The third mechanism is the resonant interaction with electromagnetic ion cyclotron waves for the rapid loss of the ring current ions (Tsurutani et al., 2018).

It is likely that unusual loss process could occur during the Carrington storm because of the uniqueness of nature. Careful diagnosis is still needed to account for the rapid recovery of the H-component of the magnetic field at Colaba. For better understanding of the rapid recovery, the equatorward boundary of the auroral emission region, as we have estimated in this paper, provides valuable information.

## 7 Conclusion

In this study, we revisited historical records in magnetically low-latitude areas being less than 35° MLAT during the stormy interval between August 28 and September 4, 1859 including the Carrington storm on September 1/2. We revisited these records in the East Asian, South American, and North American sectors, and we extracted information on their elevation angle, color, direction, and duration. Some of the low-latitude auroral displays during this storm were reddish, suggesting the possibilities of the auroras and the SAR arcs. However, the auroral displays were not only purely red. Some records include bluish, yellowish, and whitish in color. These historical records show us the equatorward boundary of auroral emission with multiple colors during the Carrington storm was 36.5° ILAT on August 28/29 and 32.7° ILAT on September 1/2. This may correspond to the equatorward boundary of the auroral oval. The equatorward boundary of purely red emission was 30.8° ILAT on September 1/2. This may correspond to the SAR arcs. The SAR arcs are not explicitly distinguished from the usual aurora caused by energetic electron precipitation without spectroscopic instruments. In that sense, we cannot definitely conclude that all the purely red emission corresponds to the SAR arcs. This reconstruction with observed auroral emission region provides conservative estimation for the intensity of magnetic storm, while findings further auroral reports suggesting extension to lower magnetic latitude could always improve this estimation. These brightness of these auroral displays was classified IBC Class IV. These facts suggest that high-energy electrons also precipitated, in part, in low-latitude areas during this stormy interval, in addition to precipitation of low-energy electrons. Finally, we compared their duration with contemporary magnetic observations and confirmed that multiple magnetic storms occurred during this stormy interval, and that the equatorward expansion of the auroral oval is consistent with the timing of magnetic disturbances.

**Acknowledgments**

We acknowledge the Bilateral Joint Research Projects between Japan (JSPS) and India (DST), the Supporting Program the "UCHUGAKU", Mission Research Projects of RISH (PI: H. Isobe), and SPIRITS 2017 (PI: Y. Kano) of Kyoto University. This work was also encouraged by a Grant-in-Aid from the Ministry of Education, Culture, Sports, Science and Technology of Japan, Grant Number JP15H05816 (PI: S. Yoden), JP15H03732 (PI: Y. Ebihara), JP16H03955 (PI: K. Shibata), JP15H05815 (PI: Y. Miyoshi), 18H01254 (PI: H. Isobe) and Grant-in-Aid for JSPS Research Fellow JP17J06954 (PI: H. Hayakawa). We gratefully thank E. W. Cliver and K. Shibata for preliminary reviews and valuable comments on our manuscript, D. Shiota for fruitful discussion on the possible impact of successive ICMEs on the Earth's magnetosphere, the archivists in the Izawa Library for permission of the research of Chikusai Nikki, K. Shiokawa for his advices on SAR arcs, S. Prosser and the Royal Astronomical Society for providing the manuscripts by R. Carrington, J. M. Vaquero and J. Humble for advices and helps on the original auroral records in Spain and Australia, and the
National Archives and Records Administration for helps and permission of research.

### References

Baker, D. N., et al. 2008, *Severe space weather events: Understanding societal and economic impacts*, National Academies Press, Washington DC

Baker, D. N., Li, X., Pulkkinen, A., Ngwira, C. M., Mays, M. L., Galvin, A. B., Simunac, K. D. C. 2013, *Space Weather*, 11, 10, 585-591. doi: 10.1002/swe.20097

Banks, P. M., Chappell, C. R., Nagy, A. F. 1974, J. Geophys. Res., 79, 10, 1459–1470, doi: 10.1029/JA079i010p01459.

Baumgardner, J., Wroten, J., Semeter, J., et al. 2008, Annales Geophysicae, 25, 12, 2593-2608.

Boteler, D. 2006, *Adv. Space Res*., 38, 2, 159–172.

Cannon, P. S., et al. 2013, *Extreme space weather*, Royal Academy of Engineering, London, UK.

Carrington, R. C. 1859, *Mon. Not. R. Astron. Soc.*, 20, 13–15. doi:10.1093/mnras/20.1.13

Carrington, R. C. 1863, Observations of the spots on the sun from November 9, 1853, to March 24, 1861, made at Redhill, London, Williams & Norgate.

Chamberlain, J. W. 1961, *Physics of the Aurora and Airglow*, doi: 10.1029/SP041.

Chapman, S. 1957, Nature, 179, 4549, 7. doi: 10.1038/179007a0

Chapman, S., Bartels, J. 1940, Geomagnetism, v. 1. Oxford Univ Press, New York.

Clette, F., Svalgaard, L., Vaquero, J. M., Cliver, E. W. 2014, Space Science Reviews, 186, 1-4, 35-103. doi: 10.1007/s11214-014-0074-2

Clette, F., Lefèvre, L. 2016, Solar Physics, 291, 9-10, 2629-2651. doi: 10.1007/s11207-016-1014-y

Cliver, E. W. 2006, *Adv. Space Res.*, 38, 119-129.

Cliver, E. W., L. Svalgaard 2004, *Solar Phys.*, 224, 407. doi: 10.1007/s11207-005-4980-z

Cliver, E. W., W. F. Dietrich 2013, *J. Space Weather Space Clim.*, 3, A31.

Cliver, E. W., N. C. Keer 2012, *Solar Phys.*, 280, 1–31.

Cole, K. D. 1965, J. Geophys. Res., 70, 7, 1689–1706, doi: 10.1029/JZ070i007p01689.

Cornwall, J. M., Coroniti, F. V., Thorne. R.M. 1970, J. Geophys. Res., 75:4699–4709. doi: 10.1029/JA075i025p04699

Cornwall, J. M., Coroniti, F. V., Thorne, R. M. 1971, J. Geophys. Res., 76, 19, 4428. doi: 10.1029/JA076i019p04428

Curto, J. J., Castell, J., Del Moral, F. 2016, *Journal of Space Weather and Space Climate*, 6, A23. doi: 10.1051/swsc/2016018

Daglis, I. A. 2000, Space Storm and Space Weather Hazards (Amsterdam, Kluwer Academic Press).

Daglis, I. A. 2003, EOS Transactions, 84, 22, 207-208.

Daglis, I. A. 2004, Effects of space Weather on Technology Infrastructure (Amsterdam, Kluwer Academic Press).

Daglis, I. A., Akasofu, S.-I. 2004, Recorder, 29, 9, 45-59.

Daglis, I. A. 2006, Space Science Review, 124, 183-202. doi: 10.1007/s11214-006-9104-z

Denton, M. H., M. F. Thomsen, H. Korth, S. Lynch, J. C. Zhang, M. W. Liemohn 2005, *J. Geophys. Res.*, 110, A07223.doi: 10.1029/2004JA010861.

Dessler, A. J., Parker, E. N. 1959, J Geophys Res 64:2239–2252. doi: 10.1029/JZ064i012p02239

Ebihara, Y., Ejiri, M.1998, Geophys Res Lett 25:3751–3754. doi: 10.1029/1998GL900006

Ebihara, Y., Ejiri, M. 2003, Space Science Reviews, 105, 1, 377-452. doi: 10.1023/A:1023905607888

Ebihara, Y., M.-C. Fok, S. Sazykin, M. F. Thomsen, M. R. Hairston, D. S. Evans, F. J. Rich, and M. Ejiri 2005, *J. Geophys. Res.*, 110, A09S22, doi:10.1029/2004JA010924.

Ebihara, Y., Fok, M. C., Immel, T. J., Brandt, P. C. 2011, J Geophys Res Phys 116:1–9. doi: 10.1029/2010ja016000

Echer, E., Gonzalez, W. D., Tsurutani, B. T., Gonzalez, A. L. C. 2008a, J. Geophys. Res., 113, A05221, doi:10.1029/2007JA012744.

Echer, E., Gonzalez, W. D., Tsurutani, B. T. 2008b, Geophysical Research Letters, 35, L06S03, doi:10.1029/2007GL031755

Echer, E., Tsurutani, B. T., Guarnieri, F. L. 2010, Journal of Atmospheric and Solar-Terrestrial Physics, 72, 4, 280-284. doi: 10.1016/j.jastp.2009.02.009

Echer, E., Tsurutani, B.T., Guarnieri, F.L, Kozyra, J.U., 2011, Journal of Atmospheric and Solar-Terrestrial Physics, 73, 11–12, 1330-1338

Ejiri, M., Hoffman, R., Smith, P. H. 1980, J. Geophys. Res., 85(A2), 653–663, doi: 10.1029/JA085iA02p00653.

Exbrayat, J. 1971, *Historia de Montería*, Cordoba.

Ezcurra, E., Mazari-Hiriart, M. 1996, Science and Policy for Sustainable Development, 38:1, 6-35, doi: 10.1080/00139157.1996.9930972

Farrona, A.M., Gallego, M.-C., Vaquero, J.M., Dominguez-Castro, F. 2011, Acta Geod. Geoph. Hung., 46, 3, 370-377, doi: 10.1556/AGeod.46.2011.3.7

Fujii, R., H. Fukunishi, S. Kokubun, M. Sugiura, F. Tohyama, H. Hayakawa, K. Tsuruda, and T. Okada 1992, *J. Geophys. Res.*, 97(A7), 10703–10715, doi:10.1029/92JA00171.

Gonzalez, W. D., B. T. Tsurutani, 1987, Planet. Space Sci., 35, 1101. doi: 10.1016/0032-0633(87)90015-8

Gonzalez, W. D., J. A. Joselyn, Y. Kamide, H. W. Kroehl, G. Rostoker, B. T. Tsurutani, V. M. Vasyliunas 1994, *J. Geophys. Res.*, 99, A4, 5771–5792. doi: 10.1029/93JA02867.

Gonzalez-Esparza J. A., Cuevas-Cardona, M. C. 2018, Space Weather, 16. doi: 10.1029/2017SW001789

Green, J. L., S. Boardsen, S. Odenwald, J. Humble, K. A. Pazamickas 2006, *Adv. Space Res.*, 38, 2, 145–154.

Green, J. L., S. Boardsen 2006, *Adv. Space Res.*, 38, 130–135.

Hapgood, M. A. 2011, *Adv. Space Res.*, 47, 12, 2059-2072.doi: 10.1016/j.asr.2010.02.007.

Hapgood, M. A. 2012, *Nature*, 484, 311–313. doi: 10.1038/484311a.

Hamilton, D. C., Gloeckler, G., Ipavich, F. M., Wilken, B., & Stuedemann, W. 1988, Journal of Geophysical Research, 93(A12), 14343–14355. https://doi.org/10.1029/JA093iA12p14343

Hasegawa, A., Mima, K. 1978, J. Geophys. Res., 83(A3), 1117–1123, doi: 10.1029/JA083iA03p01117.

Hayakawa, H., K. Iwahashi, H. Tamazawa, et al. 2016a, *Publ. Astron. Soc. Jpn.*, 68, 6, 99. doi: 10.1093/pasj/psw097

Hayakawa, H., Mitsuma, Y., Ebihara, Y., et al. 2016b, Earth Planets and Space, 68, 195. doi: 10.1186/s40623-016-0571-5

Hayakawa, H., Tamazawa, H., Uchiyama, Y., et al. 2017a, *Solar Phys.*, 292, 1, 12. doi: 10.1007/s11207-016-1039-2

Hayakawa, H., Y. Mitsuma, Y. Fujiwara, et al. 2017b, *Publ. Astron. Soc. Jpn.*, 69, 2, 17. doi: 10.1093/pasj/psw128

Hayakawa, H., K. Iwahashi, Y. Ebihara, et al. 2017c, *Astrophys. J. Lett.*, 850, L31.

Hayakawa, H., et al. 2018a, Astronomy & Astrophysics, 616, A177. doi: 10.1051/0004-6361/201832735

Hayakawa, H., Ebihara, Y., Willis, D. M., et al. 2018b, The Astrophysical Journal, 862, 15. doi: 10.3847/1538-4357/aaca40.

Hayakawa, H., Vaquero, J. M., Ebihara, Y. 2018c, Ann. Geophys., 36, 1153-1160, doi: 10.5194/angeo-36-1153-2018

Heis, E. 1859, *Wochenschrift für Astronomie, Meteorologie und Geographie*, v.2, Halle, H. W. Schmidt (WAMG, v.2)

Heis, E. 1860, *Wochenschrift für Astronomie, Meteorologie und Geographie*, v.3, Halle, H. W. Schmidt (WAMG, v.3)

Hikosaka, 1958, *Report on Ionospheric and Space Research Japan,* 12, 469–471

Hodgson, R. 1859, *Mon. Not. R. Astron. Soc.*, 20, 15–16. doi: 10.1093/mnras/20.1.15

Horwitz, J. L., Cobb, W. K., Baugher, C. R., et al. 1982, J. Geophys. Res., 87(A11), 9059–9069, doi: 10.1029/JA087iA11p09059.

Humble, J. E. 2006, Adv. in Space Res., 38, 2, 155-158. doi:10.1016/j.asr.2005.08.053

Hunten, D. M., Roach, F. E., Chamberlain, J. W. 1956, Journal of Atmospheric and Terrestrial Physics, 8, 6, 345-346. doi: 10.1016/0021-9169(56)90111-8

Iijima, T., and T. A. Potemra 1976, *J. Geophys. Res.*, 81(13), 2165–2174, doi:10.1029/JA081i013p02165.

Illing, R. M. E., Hundhausen, A. J. 1986, Journal of Geophysical Research, 91, 10951-10960. doi: 10.1029/JA091iA10p10951

Jackson A., A. R. T. Jonkers, M. Walker 2000, *Philos. Trans. R. Soc. London, Philos. Trans. Math. Phys. Eng. Sci.*, 358, 957

Janev, R. K. Smith, J. J. 1993, Cross sections for collision processes of hydrogen atoms with electrons, protons, and multiply-charged ions, in Atomic and Plasma-Material Interaction Data for Fusion, Int. At. Energ. Agency, 4, 78-79.

Jones, H. S. 1955, Sunspot and Geomagnetic Storm Data, Her Majesty's Stationary Office, London.

Jordanova, V. K., Farrugia, C. J., Thorne, R. M., Khazanov, G. V., Reeves, G. D., Thomsen, M. F. 2001, J Geophys Res 106:7. doi: 10.1029/2000JA002008

Kamide, Y., Yokoyama, N., Gonzalez, W., Tsurutani, B. T., Daglis, I. A., Brekke, A., Masuda, S. 1998, J. Geophys. Res., 103, A4, 6917–6921, doi:10.1029/97JA03337.

Keika, K., Ebihara, Y., and Kataoka, R., 2015, Earth, Planet. Space, 67:65, doi:10.1186/s40623-015-0234-y

Kimball, D.S. 1960, A study of the aurora of 1859. Sci. Rep. No. 6, University of Alaska, No. 6.

Kokubun, S. 1972, Planetary and Space Science, 20, 7, 1033-1049. doi: 10.1016/0032-0633(72)90214-0

Korte, M., C. Constable 2011, *Phys. Earth Planet. Inter.*, 188, 247–259. doi: 10.1016/j.pepi.2011.06.017

Kozyra, J. U., Nagy, A. F., Slater, D. W. 1997, Reviews of Geophysics, 35, 2, 155-190. doi: 10.1029/96RG03194

Kozyra, J. U., Fok, M.-C., Sanchez, E. R., Evans, D. S., Hamilton, D. C., Nagy, A. F. 1998, Journal of Geophysical Research, 103(A4), 6801. https://doi.org/10.1029/97JA03330

Kumar, S., B. Veenadhari, S. T. Ram, R. Selvakumaran, S. Mukherjee, R. Singh, B. D. Kadam 2015, *J. Geophys. Res.* 7307–7317, doi:10.1002/2015JA021661.

Lakhina, G. S., Alex, S., Tsurutani, B. T. Gonzalez, W. D. 2013, Supermagnetic Storms: Hazard to Society. In: *Extreme Events and Natural Hazards: The Complexity Perspective* (eds. A. S. Sharma, A. Bunde, V. P. Dimri and D. N. Baker). doi:10.1029/2011GM001073

Lakhina, G. S., Tsurutani, B. T. 2016, *Geosci. Lett.*, 3, 5. doi: 10.1186/s40562-016-0037-4

Lakhina, G. S., Tsurutani, B. T. 2017, Supermagnetic Storms: Past, Present, and Future, in: Extreme Events in Geospace (N. Buzulukova, ed.), pp.157-185. doi: 10.1016/B978-0-12-812700-1.00007-8

Lefèvre, L., Vennerstrøm, S., Dumbović, M., et al. 2016, *Solar Physics*, 291, 5, 1483-1531. doi: 10.1007/s11207-016-0892-3

Li, X., Temerin, M., Tsurutani, B. T., Alex, S. 2006, Advances in Space Research, 38(2), 273–279. https://doi.org/10.1016/jasr.2005.06.070

Liu, Y. D., Luhmann, J. G., Kajdič, P., et al. 2014, *Nature Communications*, 5, 3481. doi: 10.1038/ncomms4481

Lockwood, M., Owens, M., Barnard, L., Scott, C., Watt, C. Bentley, S. 2017, *Journal of Space Weather and Space Climate*, 8, A19. doi: 10.1051/swsc/2017048

Loomis, E.: 1859, *American Journal of Science*, Second Series, 29, 84: 385-408. (L1)

Loomis, E.: 1860a, *American Journal of Science*, Second Series, 29, 85: 92-97. (L2)

Loomis, E.: 1860b, *American Journal of Science*, Second Series, 29, 86: 249-265. (L3)

Loomis, E.: 1860c, *American Journal of Science*, Second Series, 30, 88: 79-94. (L4)

Loomis, E.: 1860d, *American Journal of Science*, Second Series, 30, 90: 339-361. (L5)

Loomis, E.: 1861a, *American Journal of Science*, Second Series, 32, 94: 71-84. (L6)

Loomis, E.: 1861b, *American Journal of Science*, Second Series, 32, 96: 318-331. (L7)

Loomis, E.: 1865, *Annual Report of the Smithsonian Institute*: 208-248 (L8)

Love, J. J. 2012, *Geophys. Res. Lett.*, 39, L10301, doi: 10.1029/2012GL051431.

Love, J. J. 2018, *Space Weather*, 16, 37–46. doi: 10.1002/2017SW001795

Lui, A. T. Venkatesan, Y., D., Anger, C. D., Akasofu, S.-I., Heikkila, W. J., Winningham, J. D., Burrows, J. R. 1977, J. Geophys. Res., 82(16), 2210–2226, doi: 10.1029/JA082i016p02210.

Maehara, H., Shibayama, T., Notsu, Y., et al. 2015, *Earth, Planets and Space*, 67, 59. doi: 10.1186/s40623-015-0217-z

Mannucci, A. J., Tsurutani, B. T., Iijima, B. A., et al. 2005, *Geophys. Res. Lett.*, 32, 12, L12S02. doi: 10.1029/2004GL021467

Mayaud, P.N. 1980, *Derivation, Meaning, and Use of Geomagnetic Indices*, Washington DC, AGU Publication, doi: 10.1029/GM022

Mendillo, M., Baumgardner, J. & Wroten, J., 2016, *Journal of Geophysical Research: Space Physics*, 121(1), pp.245–262. doi:/10.1002/2015JA021722.

Miyaoka, H., Hirasawa, T., Yumoto, K., and Tanaka, Y., 1990, Proc. Japan Acad., 66, Ser. B., pp. 47-51, doi: 10.2183/pjab.66.47

Moos, N.A.F. 1910a, *Magnetic observations made at the Government Observatory for the period 1846 to 1905 and their discussion by N.A.F. Moos. Part I. Magnetic data and instruments*, Bombay.

Moos, N.A.F. 1910b, *Magnetic observations made at the Government Observatory for the period 1846 to 1905 and their discussion by N.A.F. Moos. Part II. The phenomenon and its discussions*, Bombay.

Moreno Cárdenas, F., Cristancho Sánchez, S., Vargas Domínguez, S. 2016, Advances in Space Research, 57, 1, 257-267. doi: 10.0.3.248/j.asr.2015.08.026

Muller, C. 2014, *Orig. Life Evol. Biosph.*, 44, 3, 185–195. doi: 10.1007/s11084-014-9368-3

Neumeyer, G. 1864, *Meteorological and Nautical Taken in the Colony of Victoria*, Melbourne, Ferres.

Nevanlinna, H. 2004, Annales Geophysicae, 22, 5, 1691-1704. doi: 10.5194/angeo-22-1691-2004

Nevanlinna, H. 2006, *Advances in Space Research*, 38, 2, 180-187. doi: 10.1016/j.asr.2005.07.076

Nevanlinna, H. 2008, *Adv. Space Res.*, 42, 171–180.doi: 10.1016/j.asr.2005.07.076

Ngwira, C. M., Pulkkinen, A., Kuznetsova, M. M., Glocer, A. 2014, *J. Geophys. Res. Space Phys.*, 119, 4456–4474, doi:10.1002/2013JA019661.

Odenwald, S. 2007, *Space Weather*, 5, 11, S11005. doi: 10.1029/2007SW000344

Odenwald, S. 2015, *Solar Storms: 2000 years of human calamity*, San Bernardino, Create Space Independent Publishing Platform.

Oguti, T. 1975, Metamorphoses of aurora, Memoirs of National Institute of Polar Research. Series A, Aeronomy (ISSN 0386-5517) No.12, https://nipr.repo.nii.ac.jp/?action=repository_uri&item_id=447&file_id=18&file_no=1

Onda, K., Itikawa, Y. 1995, Proceedings of the NIPR Symposium on Upper Atmosphere Physics 8, 24-36. https://ci.nii.ac.jp/els/contents110000029809.pdf?id=ART0000355395

Oughton, E., et al. 2016, *Helios Solar Storm Scenario*, Centre for Risk Studies, University of Cambridge.

Phaneuf, R. A., Janev, R.K., and Pindzola, M. S.: 1987, Atomic Data for Fusion, Vol.V, Collisions of Carbon and Oxygen Ions with Electrons, H, H2 and H2, Tech. Rep. ORNL-6090/V5, Oak Ridge Natl. Lab., Oak Ridge, Tenn.

Rairden, R. L., Frank, L. A., & Craven, J. D. 1986, Journal of Geophysical Research, 91(A12), 13613. https://doi.org/10.1029/JA091iA12p13613

Rees, M. H., Roble, R. G. 1975, Rev. Geophys., 13(1), 201–242, doi: 10.1029/RG013i001p00201.

Rees, M. H., 1989, Physics and Chemistry of the Upper Atmosphere, Cambridge and New York, Cambridge University Press.

Riley, P. 2012, Space Weather, 10, 2, 02012. doi: 10.1029/2011SW000734

Riley, P., Love, J. J. 2017, *Space Weather*, 15, 1, 53-64. doi: 10.1002/2016SW001470

Riley, P., Baker, D., Liu, Y.D., Verronen, P., Singer, H., Güdel, M. 2018, *Space Sci. Rev.*, 214, 21. doi: 10.1007/s11214-017-0456-3

Roach, F. E., Roach, J. R. 1963, Planetary and Space Science, 11, 5, 523-540. doi: 10.1016/0032-0633(63)90076-X

Schrijver, C. J., Beer, J., Baltensperger, U., et al. 2012, *J. Geophys. Res.*, 117, A8, A08103.doi: 10.1029/2012JA017706

Schwenn, R. 2006, *Living Rev. Sol. Phys.***3**, 2., doi: 10.12942/lrsp-2006-2

Shibata, K., Isobe, H., Hillier, A., et al. 2013, *Publications of the Astronomical Society of Japan*, 65, 3, 49. doi: 10.1093/pasj/65.3.49

Shiokawa, K., Meng, C.-I., Reeves, G. D., Rich, F. J., Yumoto, K. 1997, Journal of Geophysical Research, 102, A7, 14237-14254. doi: 10.1029/97JA00741

Shiokawa, K., Anderson, R. R., Daglis, I. A., Hughes, W. J., Wygant, J. R. 1999, Phys. Chem. Earth, 24, 281-285.

Shiokawa, K., T. Ogawa, Y. Kamide 2005, *J. Geophys. Res.*, *110* (A5), 1–15, doi:10.1029/2004JA010706.

Shiota, D., Kataoka, R. 2016, *Space Weather*, 14, 56–75, doi: 10.1002/2015SW001308.

Silverman, S. M. 1995, *J. Atmos. Terr. Phys.*, 57, 6, 673 − 685. Doi: 10.1016/0021-9169(94)E0012-C

Silverman, S. M., E. W. Cliver 2001, *J. Atmos. Sol.-Terr. Phys.*, 63, 5, 523–535. Doi: 10.1016/S1364-6826(00)00174-7

Silverman, S. M. 2003, *J. Geophys. Res.*, 108, 8011, A4, doi: 10.1029/2002JA009335

Silverman, S. M. 2006, *Adv. Space Res.*, 38, 2, 136–144. Doi: 10.1016/j.asr.2005.03.157

Silverman, S. M. 2008, *J. Atmos. Sol.-Terr. Phys.*, 70, 10, 1301–-1308. Doi: 10.1016/j.jastp.2008.03.012

Smith, P. H., R. A. Hoffman 1974, J. Geophys. Res., 79, 966–967, Doi: 10.1029/JA079i007p00966

Solomon, S. C., Hays, P. B., Abreu, V. J. 1988, J. Geophys. Res., 93, A9, 9867–9882, doi: 10.1029/JA093iA09p09867.

Størmer, C. 1955, *The polar aurora*, Oxford, Oxford University Press

Svalgaard, L., Schatten, K. H. 2016, Solar Physics, 291, 9-10, 2653-2684. doi: 10.1007/s11207-015-0815-8

Takahashi, T., Shibata, K. 2017, The Astrophysical Journal, 837, L17. doi: 10.3847/2041-8213/aa624c

Tappin, S. J. 2006, *Solar Physics*, 233, 233-248, doi:10.1007/s11207-006-2065-2

Tinsley, B. A., R. P. Rohrbaugh, H. Rassoul, et al. 1984, *Geophys. Res. Lett.*, 11, 572-575. doi: 10.1029/GL011i006p00572

Tinsley, B. A., R. Rohrbaugh, H. Rassoul, Y. Sahai, N. R. Teixeira, D. Slater 1986, *J. Geophys. Res.*, 91, A10.

Tsurutani, B. T., Gonzalez, W. D., Tang, F., Akasofu, S. I., Smith, E. J. 1988, J. Geophys. Res., 93, A8, 8519–8531, doi:10.1029/JA093iA08p08519.

Tsurutani, B. T., Gonzalez, W. D., Tang, F., Lee, Y. T. 1992, Geophysical Research Letters, 19, 73-76

Tsurutani, B. T., Lakhina, G. S. 1997, Reviews of Geophysics, 35, 4, 491.

Tsurutani, B. T., W. D. Gonzalez, G. S. Lakhina, S. Alex 2003, *J. Geophys. Res.*, 108, 1268, A7.doi:10.1029/2002JA009504.

Tsurutani, B. T., Gonzalez, W. D., Lakhina, G. S., Alex, S. 2005, *J. Geophys. Res.*, 110, A09227, doi: 10.1029/2005JA011121.

Tsurutani, B. T., Verkhoglyadova, O. P., Mannucci, A. J., et al. 2007, *J. Geophys. Res.*, 113, A5, A05311.doi: 10.1029/2007JA012879

Tsurutani, B. T., E. Echer, F. L. Guarnieri, J. U. Kozyra 2008, *Geophys. Res. Lett.*, 35, 6, L06S05. doi: 10.1029/2007GL031473.

Tsurutani, B. T., Lakhina, G. S. 2014, Geophys. Res. Lett., 41, 287–292, doi:10.1002/2013GL058825.

Tsurutani, B. T., Lakhina, G. S., Echer, E., Hajra, R., Nayak, C., Mannucci, A. J., & Meng, X. 2018, Journal of Geophysical Research: Space Physics, 123, 1388–1392. doi: 10.1002/2017JA024779

Usoskin, I. G., Kovaltsov, G. A. 2012, *The Astrophysical Journal*, 757, 1, 92. doi: 10.1088/0004-637X/757/1/92

Usoskin, I. G. 2017, *Living Rev. Sol. Phys.*, 14, 3. doi: 10.1007/s41116-017-0006-9

Vaquero, J.M., Trigo, R.M. Gallego, M.C. *Earth Planet Space*, 59, e49. doi: 10.1186/BF03352061

Vaquero, J. M., M. A. Valente, R. M. Trigo, P. Ribeiro, M. C. Gallego 2008, *J. Geophys. Res.*, 113, A8, A08230, doi: 10.1029/2007JA012943

Vaquero, J. M., M. Vázquez 2009, *The Sun Recorded Through History: Scientific Data Extracted from Historical Documents*, Berlin: Springer.

Vaquero, J. M., Gallego, M. C., Dominguez-Castro, F. 2013, Geofísica Internacional, 52, 1, 87. doi: 10.1016/S0016-7169(13)71464-8

Vasyliunas, V. M., Low energy particle fluxes in the geomagnetic tail, Polar Ionosphere and Magnetospheric Processes G. Skovli, 25–47, Gordon and Breach, New York, 1970.

Viljanen, A., Myllys, M., Nevanlinna, H. 2014, *Journal of Space Weather and Space Climate*, 4, A11. doi: 10.1051/swsc/2014008

Wentworth, R. C., MacDonald, W. M., Singer, S. F. 1959, Phys Fluids 2:499. doi: 10.1063/1.1705940

Willis, D. M., F. R. Stephenson, J. R. Singh 1996, *QJRAS*, 37, 733-742.

Willis, D. M., Stevens, P. R., Crothers, S. R. 1997, Annales Geophysicae, 15, 719. doi: 10.1007/s00585-997-0719-5

Willis, D. M., F R Stephenson 2001, *Ann. Geophys.*, 19, 3, 289-302. doi: 10.5194/angeo-19-289-2001.

Willis, D. M., G. M. Armstrong, C. E. Ault, F. R. Stephenson 2005, *Ann. Geophys.*, 23, 3, 945-971. doi: 10.5194/angeo-23-945-2005

Willis D.M., Stephenson F.R., Fang H. 2007, Annales Geophysicae, 25, 417-436.

Yashiro, S., Gopalswamy, N., Michalek, G., St. Cyr, O. C., Plunkett, S. P., Rich, N. B., Howard, R. A. 2004, J. Geophys. Res., 109, A07105, doi: 10.1029/2003JA010282.

Yermolaev, Y. I., I. G. Lodkina, N. S. Nikolaeva, M. Y. Yermolaev 2013, J. Geophys. Res. Space Physics, 118, 4760–4765, doi: 10.1002/jgra.50467.

Yokoyama, N., Kamide, Y., Miyaoka, H. 1998, Annales Geophysicae, 16, 566, https://doi.org/10.1007/s00585-998-0566-z

**Appendix 1: References of Historical Documents**

In Text S1, We provide the references of historical sources for the Table 1. The abbreviations used in "Ref" show where these records are from: L (E. Loomis's publications in American Journal of Science (AJS)), RG24 (Record Group 24 of the National Archives of the United States: Logs of the U. S. Naval Ships and Stations, 1801-1946), HC (Historical Records in China), HJ (Historical Records in Japan), and MX (Mexican Newspapers). When they are published as books, journal articles, or critical editions, we provide their publication name, volume, and page number. When they are unpublished manuscripts, we provide their title, volume, folio number, shelf mark, and name of holding archive in their original languages, in term of traceability to original source documents. Note that full records of HC, HJ (except for HJ5) and MX have been transcribed and translated in Hayakawa et al. (2016) and Gonzalez-Esparza and Cuevas-Cardona (2018).

**Appendix 1.1. E. Loomis's notes in American Journal of Science (AJS)**

L1-8: AJS, v.28, 84, pp.403-404 (report from the same observer in WAMG (v.2, pp.385-386))

L1-8: AJS, v.28, 84, pp.404-406 (report from the same observer in WAMG (v.2, pp.386-387))

L3-27: AJS, v.29, 86, p.264

L3-28: AJS, v.29, 86, pp.264-265

L3-29: AJS, v.29, 86, p.265

L3-29: AJS, v.29, 86, p.265

L3-30: AJS, v.29, 86, p.265

L3-31: AJS, v.29, 86, pp.265-266

L3-31: AJS, v.29, 86, p.266

L4-14: AJS, v.29, 87, pp.398-399

L4-15: AJS, v.29, 87, p.399 (report from the same observer in WAMG (v.3, p.38))

L4-15: AJS, v.29, 87, p.399 (report from the same observer in WAMG (v.3, p.38))

L5-13: AJS, v.30, 88, p.88

L5-14: AJS, v.30, 88, p.88

L5-15: AJS, v.30, 88, pp.88-89

L6-2-43: AJS, v.30, 90, p.361.

L7-4: AJS, v.32, 94, p.76

L7-4: AJS, v.32, 94, p.77

## Appendix 1.2: Record Group 24 of the The National Archives and Records Administration of the United States: Logs of the U. S. Naval Ships and Stations, 1801-1946

RG24-1: Saranac, v.9/40, 1859-08-29, 18W04 9/23/01, RG24

RG24-2: Sabine, v.1/27, 1859-09-02, 18W04 9/18/03, RG24

RG24-3: St. Mary's, v.14/32, 1859-09-02, 18W04 9/20/04 RG24

## Appendix 1.3: Historical Records in China and Japan

HC1: *Luánchéngxiànzhì*, v.3, f.19b = 張惇德, 欒城縣志, 1872−74

HJ1: *Kotei Nendaiki*, p.1216 = 校定年代記, 新宮市史, 1937

HJ2: Yorioka Ubei Shojihikae, p.796 = 依岡宇兵衛諸事控, 印南町史, 1987.

HJ3-1: Yamaichi Kanagiya Matasaburo Nikki, Ansei06-08-06 = YK215-19-15, 弘前市図書館

HJ3-2: Yamaichi Kanagiya Matasaburo Nikki, Ansei06-08-06 = YK215-19-15, 弘前市図書館

HJ3-3: Yamaichi Kanagiya Matasaburo Nikki, Ansei06-08-06 = YK215-19-15, 弘前市図書館

HJ4: *Kenbun Nennen Tebikae*, p.315 = 見聞年々手控, 平鹿町郷土誌, 1969

HJ5: Chikusai Nikki, v.51, f.26a = XIV 78, 射和文庫

## Appendix 1.4: Historical Newspapers from Mexico

MX1: *La Sociedad*, 1859-09-03, p.3, col. 3

MX2: *La Sociedad*, 1859-09-12, p.2, col. 2

MX3: *La Sociedad*, 1859-09-17, p.2, col. 4

MX4: *La Sociedad*, 1859-09-03, p.3, col. 3

MX5: *La Sociedad*, 1859-10-24, p.1, col. 1

**Appendix 1.5: Historical Reports from *Wochenschrift für Astronomie, Meteorologie und Geographie* (WAMG).**
WA1: WAMG, v.3, p.270
WA2: WAMG, v.3, p.38
WA3: WAMG, v.3, p.16
WA4: WAMG, v.2, p.388

**Appendix 1.6: The original observational logbooks by Carrington MSS Carrington**

C1. MSS Carrington 1: Sunspot observations, v.2; The Royal Astronomical Society.

C3. MSS Carrington 3: Drawings of sunspots, showing the whole of the Sun's disk, v.2; The Royal Astronomical Society

**Appendix 2: Transcription of ship log observations during the Carrington Event**

In Appendix 2, we provide transcriptions and translations of historical documents whose whole text was not available in the community of space weather. In Text S2.1, we provide translation of ship log observations (RG24-1~3). In Text S2.2, we provide transcription and translation of Chikusai Nikki (HJ5). Their images are reproduced in Figures A1 and A2.

**Appendix 2.1. USS. Saranac (RG24-1)**

29[th] day of August, 1859.    Panama

0400

During the watch to the N & E was seen an aurora borealis, brilliantly red.

**Appendix 2.2. USS. Sabine (RG24-2)**

2[nd] day of September, 1859 At anchor off Greytown, Friday

At 1230 from about N by E to N.W. + upward to an altitude of about 35° began to assume a vermillion color, which gradually increased in intensity until about 130 when it had become of a deep rosy red & so continued until about 3, when its color had somewhat faded & finally became obscured by clouds.    During the continuance of the phenomenon, heavy masses of cum. clouds were flooding over that part of the heavens which, as they passed, entirely obscured the appearance, except where they were broken or detached; and stars of the 1[st], 2[nd], and even of the 3[rd] magnitude were distinctly visible through the light.

### Appendix 2.3. USS. St. Mary's (RG24-3)

2[nd] day of September, 1859. At Sea

At 12, discovered a very bright red light, bearing due north and extending over an arc of the horizon of about 40º with an altitude of about 20º.

### Appendix 3: Transcription and Translation of Chikusai Nikki (HJ5)

### Appendix 3.1. Transcription

六日夜五ツ過ゟ北ゟ少東　秋葉金比羅山之間　火光天ニ耀終夜明かた迄火光見へ候由　四日市桑名辺大火ニ哉と申唱候八日迄様子不訳　山田ゟハ京大坂なとニハあらすやなと申越候　右後日ニ及候へ共火事之沙汰無之全天変也　宮ゟ乗船桑名へ来候もの岐阜辺大火トノ風説也　何方も同様北方へ火光ヲ見ル

### Appendix 3.1. Translation

During night on 6[th] (Sep. 2, 1859), from around 20:00, at a little eastward from north, within the mountain of Akiba Kompira, fiery light shone in the heaven and fiery light was seen for whole night until the dawn. We discussed if Yokkaichi or Kuwana is in conflagration. However until the 8[th] (4 Sep. 1859), it was not attested. From Yamada, it was rumored (conflagration broke out) in Osaka. Even later, there was no conflagration reported and everything seems a celestial omen. Those who came from Miya to Kuwana also rumored that somewhere near Gifu was in conflagration. In any places, the fiery light was seen in the northward.

## Appendix 1: Reproduction of ship log observations during the Carrington Event

## Figure A1. USS. Saranac (RG24-1)

(Courtesy of the National Archives and Records Administration)

## Figure A1.2. USS. Sabine (RG24-2)

(Courtesy of the National Archives and Records Administration)

**Figure A1.3. USS. St. Mary's (RG24-3)**

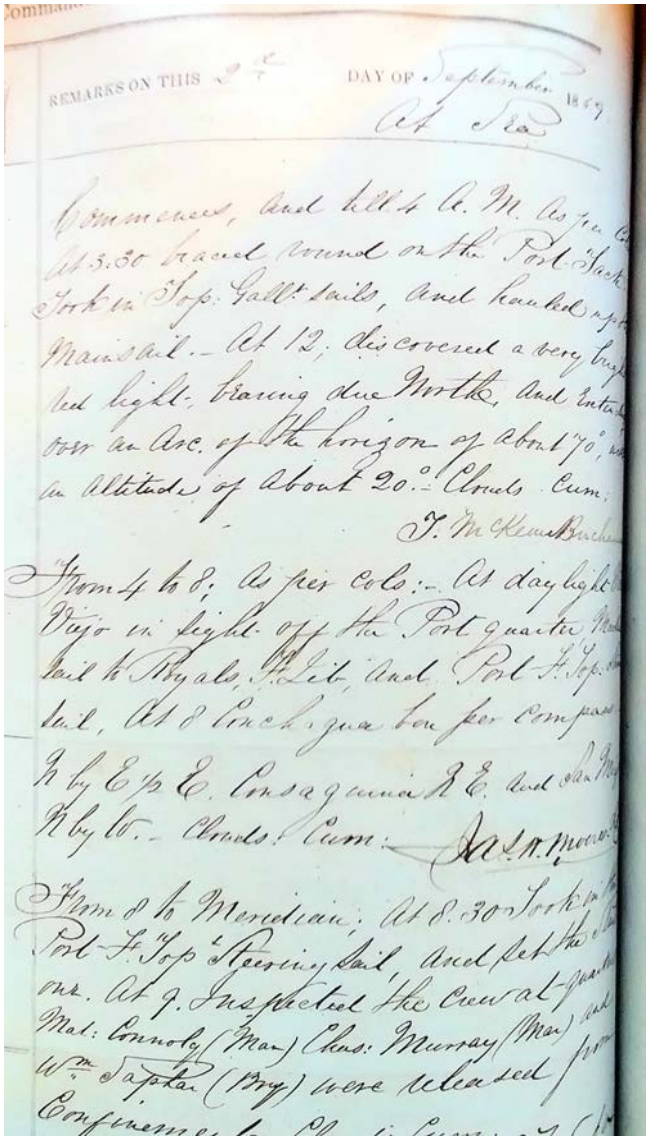

(Courtesy of the National Archives and Records Administration)

**Figure A2: Reproduction of Chikusai Nikki**

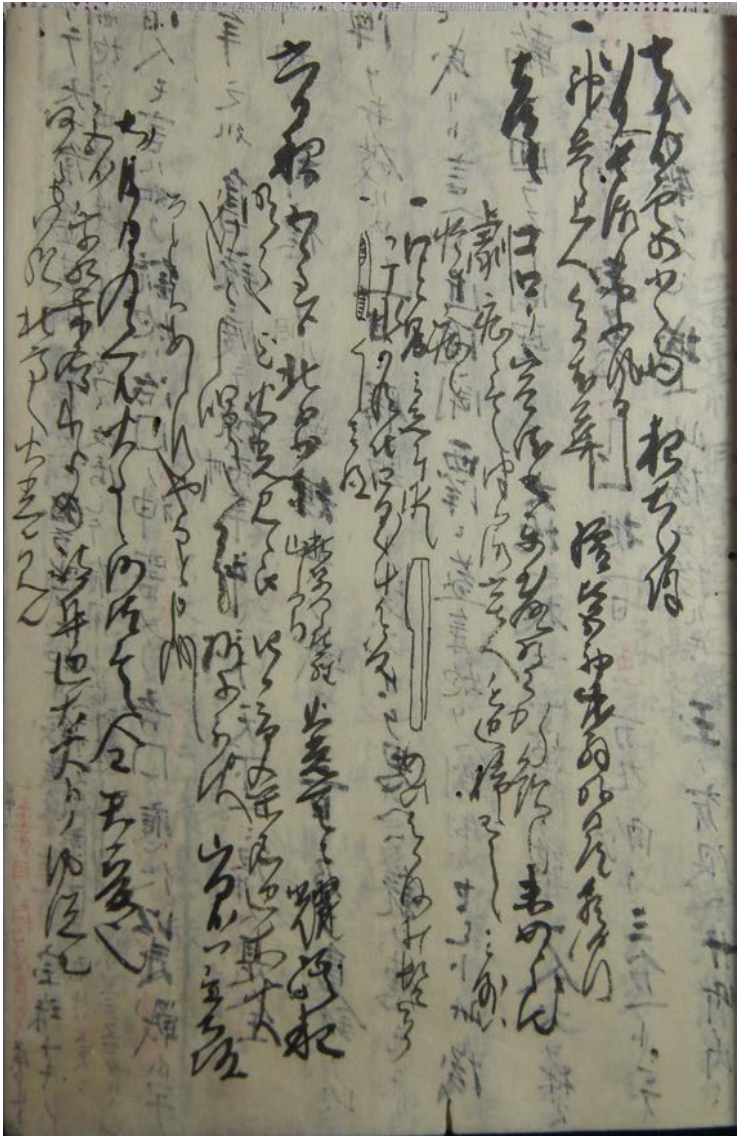

Reproduction of *Chikusai Nikki* (HJ5; Courtesy of Izawa Library). This event seems to have attracted Chikusai's interest considerable and its summary is found in the front cover of v.51 of *Chikusai Nikki*.

**Table 1.** List of Observational Sites in Magnetic Latitude below 35° MLAT. The directions are given in the points of compass: N (north), S (south), E (east), W (west), and their combination. The colors are given as R (red), W (white), P (pink), B (blue), Pu (purple), and their combination. The time is given in local time of given observational sites. In order to categorize the timing of start and end of auroral displays as a contiguous record, we define these observational dates between 06:00 and 30:00 (06:00 on the following day). "MLAT" stands for the magnetic latitude at the site. "MLAT (EB)" means the magnetic latitude of the equatorward boundary of the auroral oval at 400 km altitude, which is estimated from the information about the elevation angle. "ILAT (EB)" means the invariant latitude of the equatorward boundary of the auroral oval.